\documentclass[preprint2]{aastex}

\slugcomment{manuscript for AJ, vers. 090922}

\shorttitle{UCAC3 release paper}
\shortauthors{Zacharias et al.}

\begin{document}


\title{The Third US Naval Observatory CCD Astrograph Catalog (UCAC3)}



\author{N. Zacharias$^1$,
        C. Finch$^1$,
        T. Girard$^2$,
        N. Hambly$^3$,
        G. Wycoff$^1$,
        M.~I. Zacharias$^1$$^a$}


\author{D. Castillo$^4$,
        T. Corbin$^1$$^b$,
        M. DiVittorio$^5$,
        S. Dutta$^1$$^c$,
        R. Gaume$^1$,
        S. Gauss$^1$$^b$,
        M. Germain$^6$,
        D. Hall$^1$,
        W. Hartkopf$^1$,
        D. Hsu$^1$$^c$,
        E. Holdenried$^1$$^b$,
        V. Makarov.$^1$$^a$,
        M. Martines$^7$,
        B. Mason$^1$,
        D. Monet$^5$,
        T. Rafferty$^1$$^b$,
        A. Rhodes$^5$,
        T. Siemers$^1$,
        D. Smith$^1$,
        T. Tilleman$^5$,
        S. Urban$^1$,
        G. Wieder$^1$,
        L. Winter$^1$$^a$,
        A. Young$^1$$^c$}


\email{nz@usno.navy.mil}

\affil{$^1$U.S.~Naval Observatory, Washington DC 20392; \\
       $^2$Yale University, P.O. Box 208101, New Haven, CT 06520;\\
       $^3$Scottish Universities Physics Alliance, Institute for Astronomy,
           University of Edinburgh, Royal Observatory, Blackford Hill,
           Edinburgh EH9 3HJ, Scotland, UK;\\
       $^4$Department Of Science Operations, KM 121 CH 23, San Pedro de Atacama,
           II Region-ALMA-CHILE; \\
       $^5$U.S.~Naval Observatory, 10391 W.~Naval Observatory Rd.,
           Flagstaff AZ 86001; \\
       $^6$The Boeing Company; \\
       $^7$European Southern Observatory \\}

\affil{$^a$contractor;
       $^b$retired;
       $^c$summer student}


\begin{abstract}
The third US Naval Observatory (USNO) CCD Astrograph Catalog, UCAC3
was released at the IAU General Assembly on 2009 August 10.
It is the first all-sky release in this series and contains just over 
100 million objects, about 95 million of them with proper motions,
covering about R = 8 to 16 magnitudes.
Current epoch positions are obtained from the observations with the
20 cm aperture USNO Astrograph's ``red lens", equipped with a 4k by 4k CCD.
Proper motions are derived by combining these observations with
over 140 ground- and space-based catalogs, including Hipparcos/Tycho
and the AC2000.2, as well as unpublished measures of over 5000 plates
from other astrographs.  For most of the faint stars in
the Southern Hemisphere the Yale/San Juan first epoch plates from 
the SPM program (YSJ1) form the basis for proper motions. 
These data are supplemented by all-sky Schmidt plate survey
astrometry and photometry obtained from the SuperCOSMOS project,
as well as 2MASS near-IR photometry.
Major differences of UCAC3 data as compared to UCAC2 include
a completely new raw data reduction with improved control over
systematic errors in positions, significantly improved photometry,
slightly deeper limiting magnitude, coverage of the north pole
region, greater completeness by inclusion of double stars and 
weak detections.  
This of course leads to a catalog which is not as ``clean" as UCAC2 
and problem areas are outlined for the user in this paper.
The positional accuracy of stars in UCAC3 is about 15 to 100 mas
per coordinate, depending on magnitude, while the errors in proper
motions range from 1 to 10 mas/yr depending on magnitude and
observing history, with a significant improvement over UCAC2 
achieved due to the re-reduced SPM data and inclusion of more
astrograph plate data unavailable at the time of UCAC2.
\end{abstract}


\section{Introduction}

The US Naval Observatory (USNO) operated the 8-inch (0.2 m) Twin
Astrograph from 1998 to 2004 for an all-sky astrometric survey.
About 2/3 of the sky was observed from the Cerro Tololo Inter-American
Observatory (CTIO) while the rest of the northern sky was observed
from the Naval Observatory Flagstaff Station (NOFS).
The average number of completed fields per year was a factor of 2.0
larger at CTIO than at NOFS.
A 4k by 4k CCD with 9 $\mu$m pixel size was used in a single
bandpass (579 to 643 nm) providing a flat field of view (FOV) of just 
over 1 square degree, taking advantage of only a tiny fraction of the 
FOV delivered by the optical system of the Twin Astrograph's 
``red lens.''
A 2-fold overlap pattern of fields span the entire sky.
Each field was observed with a long (about 125 sec) and a short
(about 25 sec) exposure, thus each star should appear on at least
2 different CCD exposures, and stars in the mid-magnitude range
(about 10 to 14) should have 4 images.

UCAC3 contains just over 100 million objects, most of these are stars.
It covers the magnitude range of about R = 8 to 16 (Fig.~1)
with positional precision at mean epoch ranging from 15 to 100 mas,
depending on magnitude (Fig.~2). 
Mean position errors are shown per 1/10 mag bin with stars up to
magnitude 13 excluded whenever the formal error in either one of 
the coordinates exceeds 100 mas.  For fainter stars no such
outlier exclusion was adopted, which explains the discontinuity
in Fig.~2 and also shows what effect such a restriction has on
the derived mean formal position errors.

The distribution of proper motions is shown in Fig.~3,
and the proper motions errors as a function of magnitude
are presented in Fig.~4.
The large increase of the formal proper motion errors for
stars at magnitude 8 and brighter is caused by the saturation 
of the CCD data with associated large, formal positional errors.
The weighted mean epoch of UCAC3 data for most stars is in the 
range of 1980 to 2002 (Fig.~5), depending on magnitude as
consequence of the observing history of stars and the positional
precisions at various epochs.

The released catalog is based on all
applicable, regular survey field observations, excluding the CCD
exposures taken on extragalactic link fields and most calibration
fields.  Observations of minor planets have been extracted and will
be published separately from UCAC3.
The released UCAC3 is a compiled catalog, similar to UCAC2.
No individual epoch observations are given, nor are the pixel
data publicly available at this point.

The Tycho-2 catalog \citep{tycho2} was used as reference star catalog
to obtain UCAC3 positions on the Hipparcos System \citep{hipcat},
which is the current optical realization of the International Celestial 
Reference Frame (ICRF).
Most stars in UCAC3 have proper motions which were derived from the
astrograph CCD data combined with various earlier epoch data,
including all ground-based catalogs used also for the Tycho-2 project,
unpublished new measurements of other astrograph plates, the Southern
Proper Motion (SPM) first epoch plates, and Schmidt plate data
through the SuperCOSMOS project.
A final UCAC4 release is planned which will utilize the new
reductions of the Northern Proper Motion (NPM) program, supplementing
the SPM data, which then would allow us to derive proper motions for
all UCAC stars without the use of Schmidt plate data.
This goal could not be achieved for UCAC3 due to a production
deadline and lack of time to complete the NPM work.

The 2-Micron All-Sky Survey, 2MASS \citep{2mass} was used extensively
to analyze systematic errors of UCAC3 data and to supplement the UCAC3
catalog with near IR photometry.  Optical B,R,I magnitudes were 
copied from the SuperCOSMOS source catalog (photographic photometry)
into UCAC3 for the benefit of the users.
The number of UCAC3 objects matched with various catalogs is
presented in Table 1.

For more details about the observational data and earlier reductions
the reader is referred to the UCAC1 \citep{ucac1} and UCAC2
\citep{ucac2} papers.
Contrary to those papers, which each describe one of the earlier releases
in detail, the UCAC3 effort will be documented in a series of papers.
This paper gives the introduction aiming at the user of the UCAC3 catalog,
describing the released data, limitations, and comparisons
to other catalogs.
Technical details of the reduction process will be outlined in a paper
about the new pixel processing \citep{pxred} and a 
separate paper on the astrometric reductions leading to the mean 
positions at the CCD observing epoch \citep{ared}.
Preliminary results of these were already presented at a recent
meeting \citep{uc3aas}.
The Southern Proper Motion data re-reduction will be described 
elsewhere (Girard et al.).
Papers about double stars discovered in UCAC3 and confirmed with
speckle observing, mining UCAC3 data for new
high proper motion stars, and the extragalactic
reference frame link of UCAC are in preparation.

\section{UCAC3 versus UCAC2}

Here we summarize the main differences of UCAC3 data 
as compared to the previous release, with more
details provided in the following sections.

\begin{description}
\item[Pixel reduction:]
A completely new raw data reduction was performed for UCAC3,
applying flats and improved darks resulting in a deeper limiting magnitude.

\item[Centroiding, double stars:]
New image profile model functions were used, including double star fit
models.

\item[Completeness:]
UCAC3 is all-sky with improved completeness; however, this resulted 
in more false entries than UCAC2 had.

\item[Photometry:]
UCAC3 gives vastly improved photometry from the CCD data
re-processing. 

\item[Early epoch data:]
Many more astrograph plates (see below) were scanned at USNO 
and used to derive proper motions for UCAC3 stars.
A complete re-processing of the Southern Proper Motion (SPM)
data was performed, while the rest of the sky has only SuperCOSMOS 
early epoch data for faint stars.

\end{description}

\section{CCD Data and Processing}

\subsection{Pixel Data}

All of the 4.5 TB of compressed, raw pixel data were re-processed
for the UCAC3 release.
For the first time flats were applied to the pixel data.
An improved scheme for darks was employed which resulted in lower
background noise and a slightly deeper limiting magnitude.

Extensive research to better model the observed stellar image profiles
was undertaken, 
including investigating asymmetric model functions to account for the
skewed image shapes caused by poor charge transfer efficiency of the 
UCAC detector.  
The final reductions are based on a symmetric Lorentz profile model 
which matches the observed profile better than a Gaussian profile 
with the same number of parameters.
Details will be presented in a separate paper.

New code was developed to detect blended images of double stars and
to perform least-squares image profile fits using double star models,
fitting both components at the same time.  Many such pairs, mainly
in the 2 to 10 arcsec separation range are now being handled properly.
However, many of those pairs could only be matched to a single,
blended image in earlier epoch data to derive proper motions.
Flags in the catalog indicate the level of double star processing.
Detected pairs in UCAC data were compared to existing
double star data, and samples of potential new discoveries put on
the USNO 26-inch speckle observing program.
Results will be presented in a separate paper.

Contrary to UCAC2 the issue of completeness was pushed as much as
possible for UCAC3, which thus naturally contains many more false
detections than UCAC2 did.  Even single image detections from the
CCD data were propagated into the final catalog if they match up
with any one of the other catalogs and are above a conservative
detection threshold.  For this matching the large catalogs, 2MASS and 
SuperCOSMOS were restricted to the anticipated UCAC3 limiting magnitudes
plus some margin before performing the position based match.
This avoids accidental mismatches with very faint objects. 
Unconfirmed, faint, single images from UCAC observations are not
included in the final catalog.
Overexposed stars were propagated to the final catalog for reason
of completeness.  For those stars, and other problematic images,
the image center fit often failed, which is indicated in the
number of ``used images'' in UCAC3.  If this number is zero no fit
position could be obtained, instead the provided position is
only approximate, based on the centroid (first moments) of the light
distribution in the pixel data.

\subsection{Photometry}

UCAC3 gives 2 observed magnitudes, based on the volume of the
image profile model fitted, and a true aperture photometry,
respectively.  Extinction coefficients are derived for each
exposure with respect to Tycho-2 stars adopting a linear model
with B$-$V color.  Thus a photometric zero-point was determined
for each CCD exposure and applied to the instrumental magnitudes
to arrive at our bandpass magnitudes based on the available
Tycho-2 stars in a given field.  
  An estimate of the photometric quality of
a night is made from the average extinction coefficients of
all CCD frames taken that night and compared to other nights'
results.  Magnitudes obtained from nights
flagged as non-photometric are excluded in the calculation of
a mean magnitude for each star.  If all images are excluded,
a ``best guess'' for the zero-point of the magnitude scale on
each CCD frame is made and a mean magnitude for such stars
is derived over all available CCD frames and the error of
the magnitude is set to $-1$ in those cases.   Normally,
for each individual star 2 photometric errors have been derived.
The model error is based on the S/N ratio of the images of a
star, while the scatter error is determined from the distribution
of the individual magnitudes per star from different frames. 
The larger of these is then published in the UCAC3 catalog. 

It is expected that the photometry of the UCAC3 CCD data is
vastly improved over UCAC2, which was on the 0.3 mag level.
However, no detailed investigation into the precision or 
accuracy of photometric errors in UCAC3 has been made so far.
For well exposed stars 5 to 10\% photometric accuracy is expected.
The UCAC observing program was never envisioned to provide
reliable photometry.  No photometric standard stars were
observed to derive photometric constants for any observing night,
and all UCAC observations were performed in a single bandpass.

\subsection{Positions}

Positions in UCAC3 are on the International Celestial Reference
System (ICRS) as realized by the Tycho-2 catalog, which was used
as reference star catalog in a conventional, frame-by-frame,
astrometric reduction after various corrections were applied. 
Residuals of the final reductions are shown in Figs.~6 and 7 for
the CTIO and NOFS data, respectively.  Remaining systematic errors
are on the 5 mas level.  It is possible that these are inherent
in the Tycho-2 data, see discussion below.  
The $x$ coordinate is along right ascension (RA), while $y$ is 
along declination (Dec).

Fortunately, the 2MASS observations were made at roughly the
same epoch as UCAC observations, and the 2MASS catalog was
extensively used to derive systematic error corrections in
UCAC data.
Complex look-up tables were generated empirically to correct
for purely geometric field distortions (depending only on the 
$x,y$ coordinates of stars on CCD frames) as well as coma-like terms 
involving magnitude and $x,y$ coordinates.  These types of systematic
errors can be attributed to be in the UCAC data due to the
correlation with $x,y$ pixel coordinates.
However, a pure magnitude equation (systematic positional 
error as a function of brightness) could either be in UCAC
or 2MASS data.  
Thus the overall pure magnitude equation corrections of the
UCAC data were derived from the ``flip'' calibration data alone.
These calibration fields have been observed throughout the
UCAC project with the telescope being on one side of the pier
(East or West) then on the other.
These flip observations provide pairs of CCD exposures which
are rotated by $180^{\circ}$ with respect to each other,
revealing the magnitude equation offsets independent of
external catalog data.  The assumption here of course is
that the magnitude equation stays constant over the set
of East/West exposures and other systematic errors like
coma terms have been removed.

No corrections as a function of color were applied.
Differential color refraction effects are typically below 5 mas
due to the narrow UCAC bandpass.
For a detailed discussion of the astrometric reductions
leading to UCAC3 the reader is referred to the separate
paper by Finch et al.

\section{Proper Motions}

\subsection{New data and processing}

For UCAC3 the complete set of the second 
{\em Astronomische Gesellschaft Katalog}
(AGK2) plates, taken around 1930, could be utilized from 
scans made on the StarScan machine \citep{starscan}.
This set comprises about 1950 plates taken at the Bonn and
Hamburg observatories, covering the sky north of 
$\delta \ = \ -2.5^{\circ}$ and blue magnitude range 5 to 12.
Only a partial set of the AGK2 data was available for UCAC2.
Over 1.2 million stars (see Table 1) could be measured this
time, while ``only" about 186,000 stars were measured,
reduced and published in the original AGK2/AGK3 project from
a several decade long effort, when ``computers" were humans.

In addition, a total of about 2000 Hamburg Zone Astrograph (ZA),
900 USNO Black Birch Astrograph (yellow lens, BY), and
300 Lick Astrograph (LA) plates were scanned on StarScan
and reduced with Hipparcos reference stars to provide accurate
early epoch positions for stars down to V = 14 (ZA, BY) and
V = 16 (LA).  However, all those plates together cover only
about 1/3 of the sky, targeting fields around ICRF extragalactic
sources and special fields observed for other programs.

A complete new reduction of the SPM data was performed
applying the modified StarScan \citep{starscan} pipeline
reduction code to the Precision Measuring Machine (PMM) \citep{pmm}
pixel data.  The resulting global-plate $x,y$ center coordinates
were processed by the Yale University reduction pipeline to 
correct for systematic errors as function of magnitude, utilizing 
all grating images of those data (Girard et al, in preparation).
The new processing of these data improved the proper 
motions of UCAC3 stars fainter than about 14th magnitude and
covered by the SPM2 by about a factor of 2 w.r.t.~the UCAC2 release.

Unfortunately, the corresponding Northern Proper Motion (NPM)
reductions did not progress fast enough for the UCAC3 schedule
and will be utilized at a later time for UCAC4.
The SuperCOSMOS data \citep{superc1} based on Schmidt 
survey plates provided early epoch positions to derive proper
motions of faint UCAC3 stars all-sky.
For each catalog used in the proper motion calculation an
estimated systematic error floor was added to the internal
errors (Table 2).
Individual CCD mean position errors (small number statistics
from scatter of a few observations per star) were already clipped
to a minimum of 10 mas prior to this step.
SuperCOSMOS data were not excluded in areas covered by SPM;
however, the SPM data with their significantly lower errors
dominate the proper motion solution, if available.

\subsection{High proper motion stars}

An effort has been made to tag previously known High Proper Motion
(HPM) stars in the UCAC3 catalog utilizing published proper motion
catalogs and surveys.  While we have made an effort to identify most
previously known HPM stars with $\mu \ge$ 0$\farcs$15 yr$^{-1}$, the
list is not complete.  In all but a few cases data for these known
proper motion stars were retrieved using the VizieR on-line data tool
at the Strasbourg Astronomical Data Center (CDS).  For the few cases
where data were not available through CDS, the data were retrieved
through the corresponding published literature. The proper motion data
given in the UCAC3 catalog for these previously known HPM stars come
from the catalogs themselves and are not computed as other proper
motions are in the UCAC3 catalog.

A list of HPM stars was compiled and each UCAC CCD frame searched
for possible images of those stars.
Those images were extracted from the regular pipeline processing
of UCAC3 to avoid confusion and mis-matches with other catalogs.
A total of 51,297 HPM stars could be identified in UCAC3 data;
they are identified by the {\em MPOS} numbers larger than 140 million.

The mean positions of the HPM stars are based on UCAC CCD observations;
however, no attempt was made to identify these stars in early epoch
catalogs to derive new proper motions.
More details and a list of references can be found in the readme
file of UCAC3 and the upcoming paper about new HPM stars and
common proper motion pairs found in UCAC3 data (Finch et al).

\section{CTIO - NOFS overlap region}

A region of the sky around $\delta = +20^{\circ}$ was observed
at CTIO and then repeated from NOFS within about 3 months
(see UCAC2 paper).  Based on these 1410 CCD frames separate
mean positions were generated from the data of each site,
utilizing the final version of the systematic error corrections,
which are different for location and telescope orientation.
For the CTIO observing the telescope was on the West side of
the pier, while at NOFS it was on the East.
Figure 8 shows the position differences between the CTIO and NOFS
based data as function of magnitude.
Large differences are found only for bright, overexposed stars
in the declination coordinate, as expected.  These are residual
systematic errors from bleeding columns of stars too bright
for precise UCAC astrometry.
All other systematic position differences are small, typically
5 mas, showing excellent consistency between the different data
sets of CCD observing (at vastly different zenith distances) 
and processing.
The data shown in Fig.~8 is not inconsistent with Figs.~6 and 7
because of the different area of the sky sampled.  Residuals
with respect to Tycho-2 reference stars vary with declination 
zone.  Figs.~6 and 7 show the summary over all applicable frames
taken at CTIO and NOFS, respectively, while Fig.~8 covers only
a small area in the sky.

\section{Comparisons with other Catalogs}

For the following comparisons with the UCAC3 release data,
only stars with unique, single matches to the following 
catalogs were used.  A match radius of 1.5 arcsec
was adopted for positions at the desired common match epoch,
by applying proper motions as specified below.

\subsection{SuperCOSMOS}

There are several star catalog solutions based on the same 
all-sky Schmidt plate surveys and different plate measures.
SuperCOSMOS \citep{superc3} proper motions were applied to the 
SuperCOSMOS Source Catalog positions to generate positions at the 
epoch of UCAC data.
Relatively small systematic position differences are found.
Figure 9 shows some examples of such systematic position differences
as a function of magnitude (UCAC model mag), RA and Dec.
The Schmidt plate pattern is clearly seen in the differences as a 
function of declination; however, overall typical systematic errors
are only on the 100 mas level, less than what earlier had been found
in USNO-B data, typically 200 mas, sometimes exceeding 300 mas 
\citep{ucac-nomad}.
Similar results were obtained from minor planet orbit determinations
based on UCAC2 and USNO-B reference stars \citep{chesley}, and a
re-processing of USNO-B is in progress.
The error contribution from the UCAC data is negligible for 
these comparisons, thus we see mostly the absolute position
errors of these Schmidt plate data when comparing to UCAC.
These results led to the decision to use SuperCOSMOS data to
derive proper motions of UCAC3 instead of the current USNO-B catalog.

\subsection{UCAC2}


Figures 10 to 20 illustrate the systematic differences between this
UCAC3 data release and the previous UCAC2 version regarding
magnitude, positions and proper motions.
Out of the 48.3 million entries in UCAC2, close to 1 million could
not be matched up with a UCAC3 entry uniquely.
We can't exclude the possibility that UCAC3 is actually missing
a significant number of {\em bona fide} stars; however,
equally well the majority of those not matched objects
could be invalid entries in UCAC2.  See also section 8 for
a discussion of UCAC3 problems and issues.
Due to the large volume of data the comparison was split up
between the northern and southern hemisphere.

For the stars in common we see a very small scatter but complex,
large, systematic difference in the photometry (Fig.~10).
This confirms reported errors in UCAC2 photometry of
typically 0.3 mag, which hopefully are resolved for the
better in UCAC3.

The RMS position differences between UCAC2 and UCAC3 at the
standard epoch of 2000.0 are shown in Fig.~11 and 12 for the
southern and northern hemisphere, respectively.
The floor level for the well exposed stars (10 to 14 mag)
is around 15 mas for the southern hemisphere data, consistent 
with a 1-sigma formal error of each catalog in this magnitude 
range.  The northern hemisphere data have an added RMS component
beyond 12th magnitude, which can be explained by the systematic
differences as function of magnitude as shown in Fig.~13 and 14.
There are clearly differences between the southern and northern 
hemisphere data; however, typical systematic UCAC2 $-$ UCAC3
position differences are only 5 to 15 mas for the entire
8 to 16 magnitude range. 

Similarly Figures 15 to 18 display the UCAC2 $-$ UCAC3 position 
differences as function of right ascension and declination.
The mean offset is dominated by the majority of the faint
stars from the offset as function of magnitude (see Fig.~13 and 14).
In general, differences in the south are smaller than in the north.
The parabola-shape differences for the RA component as function
of RA in the north is surprising. There is also a saw-tooth
pattern visible.  These were likely introduced into UCAC3 through
the use of SuperCOSMOS data propagating into the positions at
epoch 2000 through proper motion errors.

The large saw-tooth pattern in the declination differences
as function of Dec in the north (Fig.~18 bottom) is similarly
caused by the difference
between the previous (UCAC2) Yellow-Sky (NPM based) and
the UCAC3 SuperCOSMOS data, based on Schmidt plates.
This is likely an effect inherent in the proper motion differences
between the 2 sets of early epoch data for the faint stars.
The UCAC sky survey in the declination range from $0^{\circ}$ to 
$50^{\circ}$ was undertaken between about 2000 and 2003, going north, 
thus with increasing epoch difference relative to the standard 
epoch of 2000.0, at which this position comparison is performed.
This pattern is a combination of position errors and 
propagation from proper motion errors.

To understand this better we look at the proper motion differences
between UCAC2 and UCAC3 as shown in Figures 19 and 20 for the south
and north, respectively.
Averaging over all stars the mean magnitude is about 15.5,
at which we see almost $+5$ mas/yr (UCAC2$-$UCAC3) proper motion
difference in the RA component.
The data at $+50^{\circ}$ declination were taken around 2003 and
moved back by 3 years of proper motion to the comparison epoch 2000. 
Thus the 5 mas/yr proper motion error results in a position
offset of 15 mas with respect to data taken around 2000 
(thus near the equator).
This is close to what we are seeing in the RA difference plot 
of Fig.~18 (top).
Similarly, with opposite sign, Fig.~20 suggests a $-3$ mas/yr
difference in declination proper motion which translates into
a position offset of $+9$ mas in the 2003 data.  This is very similar
to the average increase of the declination offset as seen in
Fig.~18 (bottom), which goes from about +6 to +15 mas over
the $0^{\circ}$ to $+50^{\circ}$ Dec range plus the modulation 
of the saw-tooth pattern added on top of this average trend.

The period of this saw-tooth pattern is very close to $5^{\circ}$.
Both the NPM plate pattern as well as the second Palomar Observatory
Schmidt Survey (POSS) adopted a $5^{\circ}$ spacing between fields,
while the first POSS adopted a $6^{\circ}$ pattern.
The SuperCOSMOS data used for UCAC3 are based on the POSS plates,
while the Yellow-Sky catalog which was used for proper motions 
of faint stars in UCAC2 in the northern sky is based on NPM plates.
Looking only at the UCAC2$-$UCAC3 differences,
it is not clear which data caused the saw-tooth pattern seen 
in Fig.~17, i.e.~whether this is a new problem in UCAC3 or something
in UCAC2 has been fixed now.

\subsection{SPM2}

UCAC3 proper motions were compared with those from the SPM2
catalog \citep{spm2}.  The SPM2 contains about 321,000 stars,
individually selected and measured on the Yale University
PDS machine.  The plates cover a declination range of about 
$-$45 to $-25^{\circ}$, span 2 epochs about 25 years apart,
and reach a limiting magnitude near 18, deeper than UCAC.
About 205,000 of the SPM2 stars were uniquely matched with UCAC3.
The difference in proper motions (UCAC3$-$SPM2) for RA and Dec
(Figs.~21, 22) show only small (about 1 mas/yr) systematic differences
as a function of UCAC magnitude.
Figure 23 displays the position differences UCAC3$-$SPM2 at the SPM2 
epoch of 1991.25 by using the UCAC3 proper motions to bring the UCAC3
positions from 2000 to 1991.25.
Similarly, Fig.~24 shows the position differences at UCAC3 epoch
of 2000 when applying SPM2 proper motions to SPM2 positions.

Because both data sets share the first epoch SPM positions these
plots show mainly the difference between the CCD observations 
of UCAC and the second epoch SPM observations.
What we see is a mix between remaining systematic errors of UCAC3
and SPM2 which can not be separated out at this point.
However, these position differences are as small as can be hoped for,
about 3 to 10 mas.

Figures 25 and 26 similarly show the UCAC3$-$SPM2 position differences
as a function of declination.
Again, differences are small; however a saw-tooth pattern seems
to be present in the declination differences.
Again, the period seems to be close to $5^{\circ}$, pointing to
small residual systematic error contributions from the SPM2 data.

\subsection{PM2000}

The PM2000 catalog \citep{pm2000} was used in an external comparison
with UCAC2 and UCAC3.  The PM2000 covers the zone of about 
$9.5^{\circ}$ to $18.5^{\circ}$ declination.  
Position differences are shown in Fig.~27 for the epoch of 2000.0
which is very close (within about a year) of the CCD observations
of UCAC.
For the right ascension component, the PM2000 agrees with the UCAC2
better than with the UCAC3.  The reverse is true for the declination
component.  All systematic differences are at or below the 10 mas
level except for the RA component in the UCAC3$-$PM2000 comparison
at 16th magnitude.

\subsection{2MASS}

Figures 28 and 29 show position differences between UCAC3 and 2MASS
as function of magnitude, for the southern and northern hemisphere,
respectively.
The UCAC3 proper motions are used to bring the UCAC3 positions to
the 2MASS epoch (about 1998 to 2002) for each individual star matched 
uniquely.
Figures 30 and 31 show the RMS scatter of the corresponding data.
The issue of concern here is the large position difference at the
faint end for the southern hemisphere data.

For comparison, similar plots were generated using the UCAC2 data
(Fig.~32 to 35), which show clearly smaller differences than the
comparison of UCAC3 with 2MASS.
For both the UCAC2 and UCAC3 data sets versus 2MASS, the minimum RMS 
scatter (70 mas) occurs at around 12th magnitude, thus with
negligible error contribution from the UCAC data.
However, at the fainter magnitudes, in the 14 to beyond 16 mag
range UCAC2 agrees with 2MASS significantly better than UCAC3 does.

\subsection{Other, external checks}

A sample of about 800 stars of the Small Magellanic Cloud were
identified in UCAC2 and UCAC3.  The mean proper motion was
determined to be about +4.0 and $-$3.5 mas/yr for RA and Dec
respectively (UCAC2) and +0.4, $-$2.5 mas/yr for UCAC3 
(priv.~comm. P.~Massey).  The scatter in proper motion values
was found to be comparable between UCAC2 and UCAC3.

\section{The Catalog}

The UCAC3 data files are organized in 0.5 degree wide declination
zones, numbered from 1 to 360 beginning at the
South Celestial Pole.  Within each zone stars are sorted
by ascending right ascension.  Each 84 byte fixed length, binary
record contains all the data for a star.
The byte order is that of the native intel-type processor binary
data format.  For some computers a byte-swap might be needed.
Table 3 describes all data items for each star.
Detailed remarks are given in the readme file which comes
with every data distribution (DVD or online).
Sample access code (in Fortran) is provided as well.

While the MPOS number (last column on each data record;
MPOS stands for mean (CCD data) positions and is a running,
unique, internal star number)
mainly provides a means to identify known high proper motion stars,
the primary star identification number should be of the form 
3UCzzz-nnnnnn.
The ``3UC" is constant and indicates the UCAC3 catalog.
The 3 digit ``zzz'' number is the zone the star is in, followed
by a dash and a 6 digit number which is the record number of
the star in that zone.   Thus the official designation of
the star 42 in zone 7 would be  3UC007-000042.

Similar to UCAC2, UCAC3 is a compiled catalog giving the weighted
mean position and proper motion of stars based on all input
catalogs, including the CCD data.

UCAC3 does contain some non-stellar objects, mainly galaxies.
There is no star/galaxy separation parameter based on pixel data
in UCAC3.  However, flags are provided which indicate matches
with known non-stellar objects, from the 2MASS extended source 
catalog, the LEDA galaxy catalog, and non-stellar flags copied 
from the SuperCOSMOS and SPM data.

\section{Problems and Warnings}

\subsection{Overview}

UCAC3 is not as ``clean'' as UCAC2.  The goal here was to enhance
the completeness of the catalog, showing every possible star on
the sky detected within the accessible magnitude range.  This could only
be accomplished by allowing faint and uncertain objects to enter
the catalog as well, resulting in an increased fraction of erroneous 
entries in UCAC3.  Some specific problems are addressed below.

Users looking for reliable reference stars should check on some
of the flags and auxiliary data entries.
UCAC3 records without 2MASS match or without derived proper
motion are questionable.
Overexposed and problem stars do have some images from the CCD
data excluded, up to excluding all images, when the derived
position is entirely based on an unweighted mean of all
available detections (center of light).
Those stars certainly should be excluded from use as
reference stars, as should all those with an internal
position error exceeding some limit set by the user.

\subsection{Erroneous close doubles}

Due to a processing error some stars appear twice in UCAC3
with very similar positions (typically 0 to 200 mas separation).
This can happen for single stars (dsf flag = 0) or components
of true doubles, making them appear to be quadruple stars.
A total of 771,018 such erroneous pairs were identified among
the single stars with separation up to 2 arcsec, which would
have to have $dsf \ \ge \ 1$ if they were real doubles.
This is about 0.8 \% of the UCAC3 entries.
However, a simple exclusion of objects without 2MASS identification
gets rid of these false close components, leaving a single,
valid star.  This criterion alone removes over 99.4\% of the
problem cases, and might be advisable in general.
A total of about 2.65 million objects in UCAC3 have no 2MASS
identification, and some 0.77 million of those belong to
this single problem group alone.

\subsection{Bright star problems}

The adopted algorithm to detect and characterize double stars
from the UCAC pixel data erred on the side of completeness.
It did also pick up spurious noise near some bright stars
as ``new components".  A simple relationship was found (see also
upcoming paper by Mason et al.) to exclude those.
All objects with separations of less than 7 arcsec and
combined magnitudes (sum of 2 components) of less than 18
are likely false.  UCAC3 doubles in the remaining parameter
space showed a very high degree of confirmation with the
USNO speckle camera at the 26-inch telescope.  This test was 
limited by the capabilities of the instrument, reaching to 
about 12th magnitude for the secondary star.

In general, the long exposures of UCAC saturate around
magnitude 10, while the short exposures saturate at about
magnitude 8.  Thus for stars in the 8 to 10 mag range only
2 observations are available from the regular overlap pattern
of observed fields.  Stars brighter than 8th magnitude
generally have no fitted position at all but are kept in
UCAC3 when detected and a center-of-light approximate
position is available.  One should inspect the {\em nu1} 
value, giving the number of images from CCD data used for the
mean position.  If it is zero, the position given in
UCAC3 is only very approximate and can be off by
arcseconds.

Stars of about magnitude 6 and brighter are likely not
in UCAC3 at all.  
An updated version of the NOMAD catalog of all stars to about 
20th magnitude including all naked-eye stars is in preparation
as part of the UCAC4 effort.

\subsection{Missing stars}

A detailed comparison of UCAC3 with UCAC2 revealed close to
a million stars missing from UCAC3 which are in the UCAC2 release.
A match of those objects with 2MASS and the CMC14 catalog \citep{cmc14}
shows that at least 99\% of these seem to be legitimate stars
(G.Harald, J.Greaves, priv.~communication).
The magnitude distribution of these stars follow the general
distribution for UCAC3, indicating a ``random" effect.
These stars are also distributed all over the sky with clustering
along the galactic plane, roughly following the general distribution
of stars.
Most of these stars are no ``problem" candidates and the reason
for them not being in UCAC3 is not known at this point, however
a correlation to the processing error which resulted in double entries
(see above) is suspected.
UCAC3 is a completely new reduction independent of what
has been detected and processed in the UCAC2 pipeline reductions.

\subsection{Bogus proper motions}

Some proper motions will be bogus due to an incorrect match
of stars between 2 or more catalogs, sometimes spanning
several decades.
This is particularly a problem for the faint stars,
which rely sometimes on only 2 catalog positions, the
CCD observation at UCAC epoch and one other, deep
catalog at a much earlier epoch.
Fortunately, for both large catalogs (SPM and SuperCOSMOS)
we have proper motions available which were applied to
bring the positions of individual stars to the mean UCAC
epoch before matching.
Nevertheless some objects are expected to be 
mismatched and the resulting bogus proper motions 
could be large, contaminating any legitimate new
high proper motion stars in UCAC3.
This was found to be the case when checking a large sample
of such stars with real sky images \citep{chpm}.

\subsection{Systematic errors in proper motions}

The above detailed catalog comparisons reveal a possible
problem with UCAC3 proper motions at the faint end.
Due to the sole use of Schmidt plate data for proper motions of
faint stars in the north, and the large systematic differences of
UCAC3 w.r.t.~2MASS for faint stars in the south, we recommend using
stars of about 16th magnitude and fainter in UCAC3 with caution.
If possible, they should be avoided as reference stars.

\section{Discussions and Conclusions}

UCAC3 is the first all-sky catalog of this series.
From the details presented above it appears that the
systematic errors of the CCD observations for UCAC3 are 
corrected even better than they were in UCAC2 (see for example
the CTIO / NOFS overlap area, the comparison with SPM2 and
PM2000).
The magnitude dependent systematic errors seem
to be well controlled, with the exception of the very
faint end of UCAC3 (around 16th mag and fainter).
Comparing Figures 13, 24, 28 and 32 suggests a systematic
error in UCAC3 positions (both coordinates) as a function
of magnitude for stars around magnitude 16.  
UCAC2, SPM2 and 2MASS agree, while differences of any
of these catalogs with UCAC3 show some systematic deviations.

The use of Schmidt Survey data likely caused a problem
for the proper motions of faint stars (R $\ge$ 14), even partly
affecting the area covered by the new reductions of the SPM data,
and particularly affecting the northern hemisphere.
Although the formal errors in proper motions significantly
dropped for a large number of stars as compared to UCAC2,
systematic errors as function of location on Schmidt plates
crept into the UCAC3 catalog, increasing the scatter when
compared for example with the 2MASS catalog.

A significant improvement of the photometry in UCAC3 was 
achieved, which is handled properly for the first time.
The complete re-reduction of the pixel data also extended
the limiting magnitude, providing more and fainter stars
in UCAC3 than in earlier releases.

At the bright end,
the residuals of the Tycho-2 reference stars show some remaining
magnitude equations.  If we assume the internal calibrations of
the UCAC3 CCD observations (utilizing the East/West flip data)
are correct this indicates magnitude equations in the Tycho-2
catalog itself of about 1 to 2 mas/mag.
Assuming the Tycho space-based observations are free of such
errors, this indicates uncorrected errors in the order of
100 to 200 mas in the Astrographic Catalog (AC) whose average
epoch is around 1900.
The AC is the major ground-based catalog used to obtain the
Tycho-2 proper motions.

The major remaining steps to be taken to conclude this project are
a) utilize overlap conditions of the regular 2-fold center-in-corner
pattern of fields observed with the CCD astrograph to reduce 
coordinate dependent errors introduced by the reference stars,
b) include re-reductions of the NPM data for proper motions in
the north, eliminating the need to resort to Schmidt plate data,
c) check on the extragalactic link by employing the dedicated
observations in ICRF fields and their corresponding deep CCD
imaging with larger telescopes, and d) fix above mentioned
problems.
These are the goals for UCAC4.


\acknowledgments

We are grateful to the additional observers S.~Pizarro (CTIO),
S.~Potter and D.~Marcello (NOFS), as well as the entire staff
at CTIO and NOFS who at one time or another assisted with this program,
in particular O.~Saa and M.~Smith (CTIO), B.~Canzian and C.~Dahn (NOFS).
We thank the USNO Washington and NOFS instrument shops, in particular
J.~Pohlman for support, as well as G.~Hennessy for system administration
support.
K.~Seidelmann is thanked for his role in getting this project going.
Spectral Instruments, in particular G.~Sims is thanked for the 
outstanding support of our camera. 
B.~Gray is thanked for customizing his {\em Guide} software for
UCAC needs.
National Optical Astronomy Observatories (NOAO) are acknowledged for
IRAF, Smithonian Astrophysical Observatory for DS9 image display software,
and the California Institute of Technology for the {\em pgplot} software.
More information about this project is available at \\
\url{http://www.usno.navy.mil/usno/astrometry/}.





\clearpage

\begin{figure}
\epsscale{1.00}
\plottwo{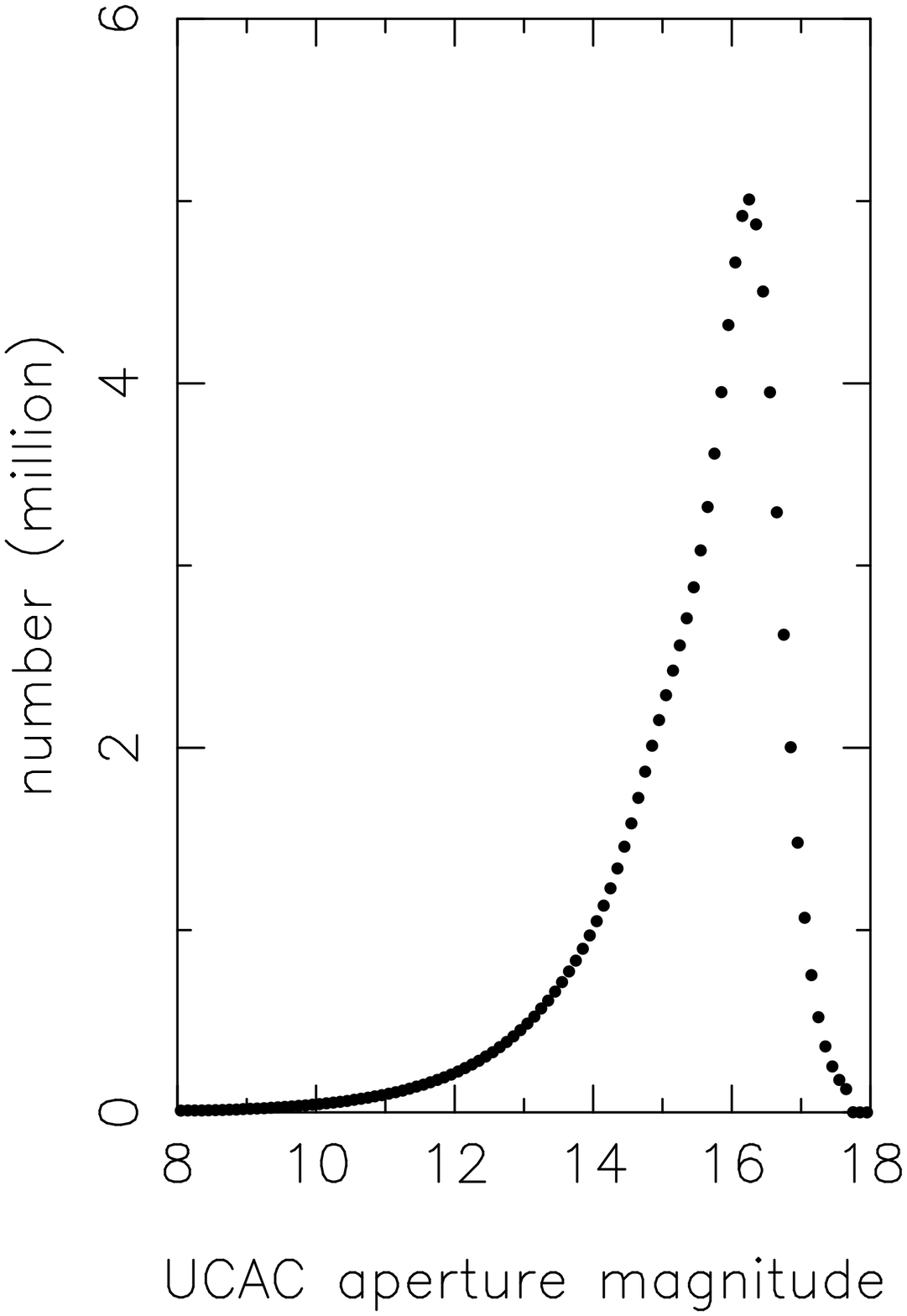}{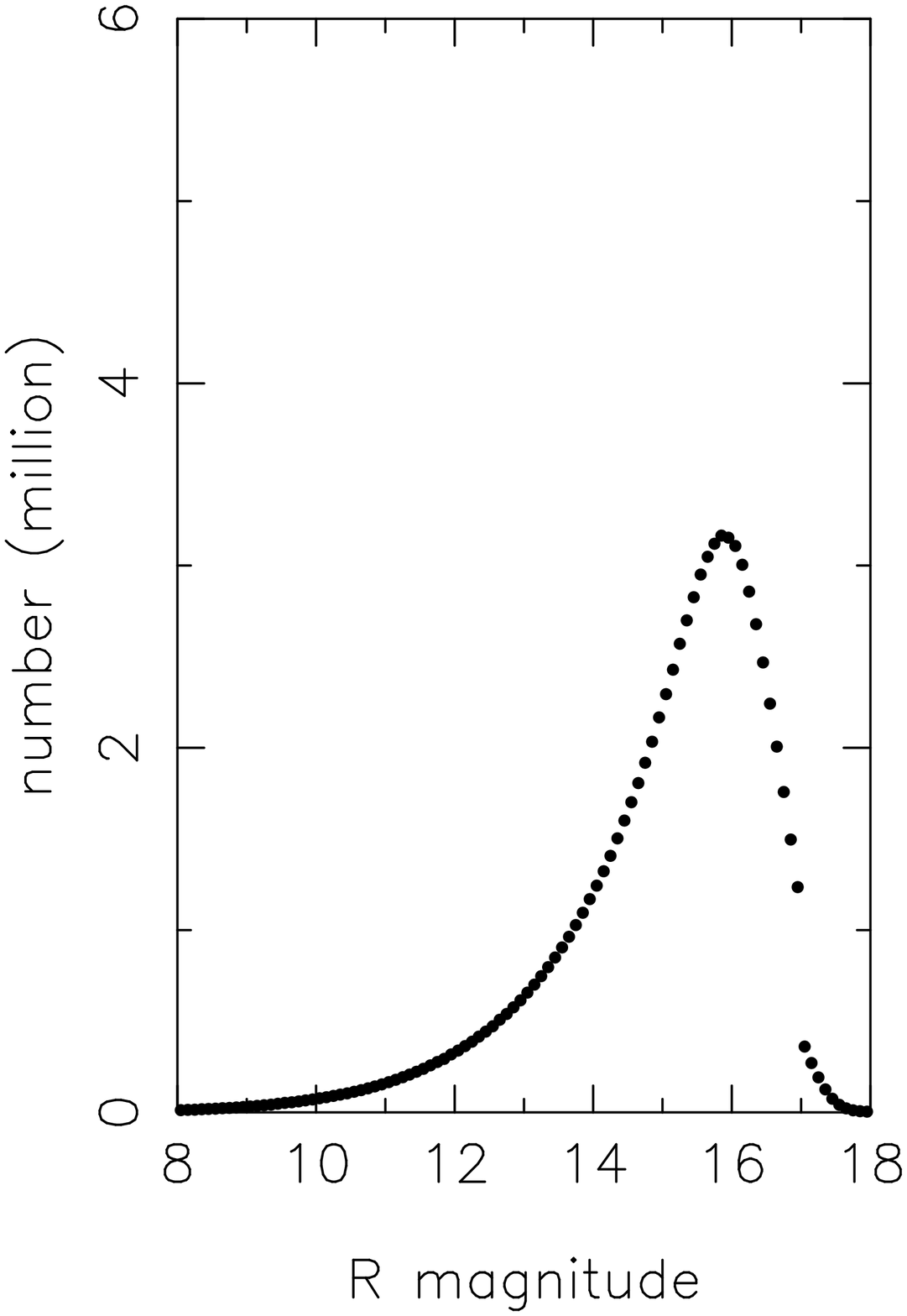}
\caption{Distribution of UCAC3 stars as function of UCAC3 aperture
 magnitude (left) and SuperCOSMOS R magnitude (right).  
 The limiting magnitude is close to 16.0 in both cases.}
\end{figure}


\begin{figure}
\epsscale{1.00}
\plotone{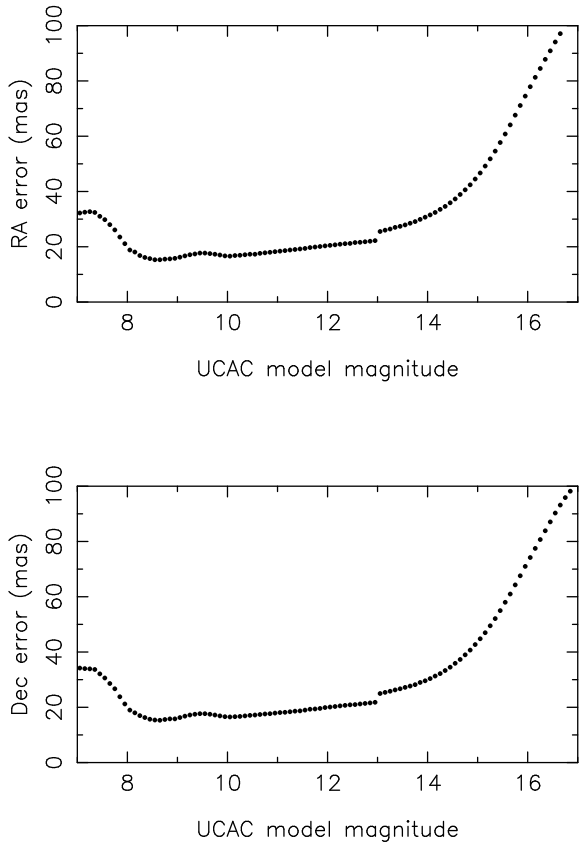}
\caption{UCAC3 formal, mean position errors per coordinate at central 
  epoch as function of model fit magnitude.
  The discontinuity at magnitude 13 is artificial due to
  the adopted outlier exclusion, see text.}
\end{figure}


\begin{figure}
\epsscale{1.00}
\plottwo{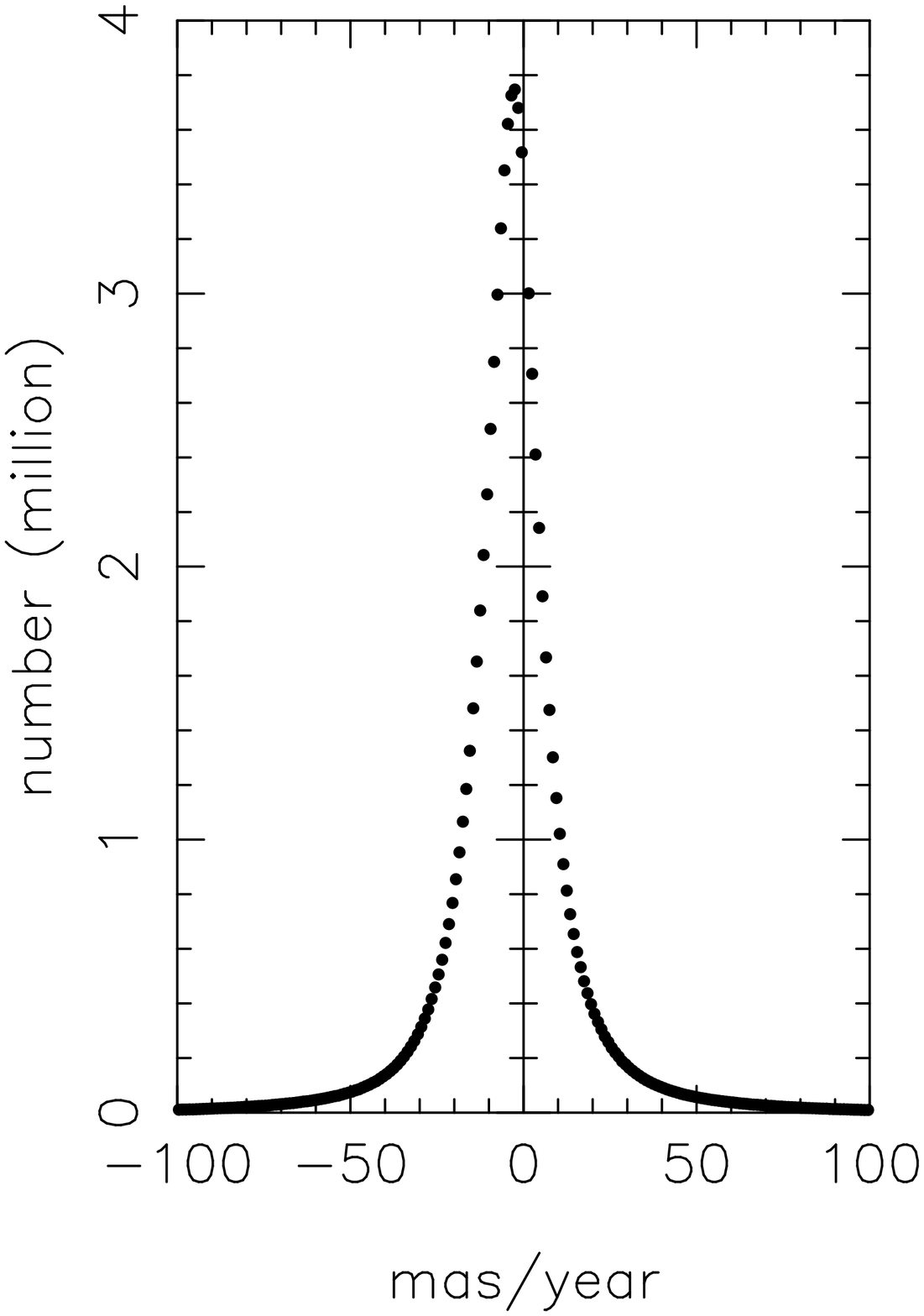}{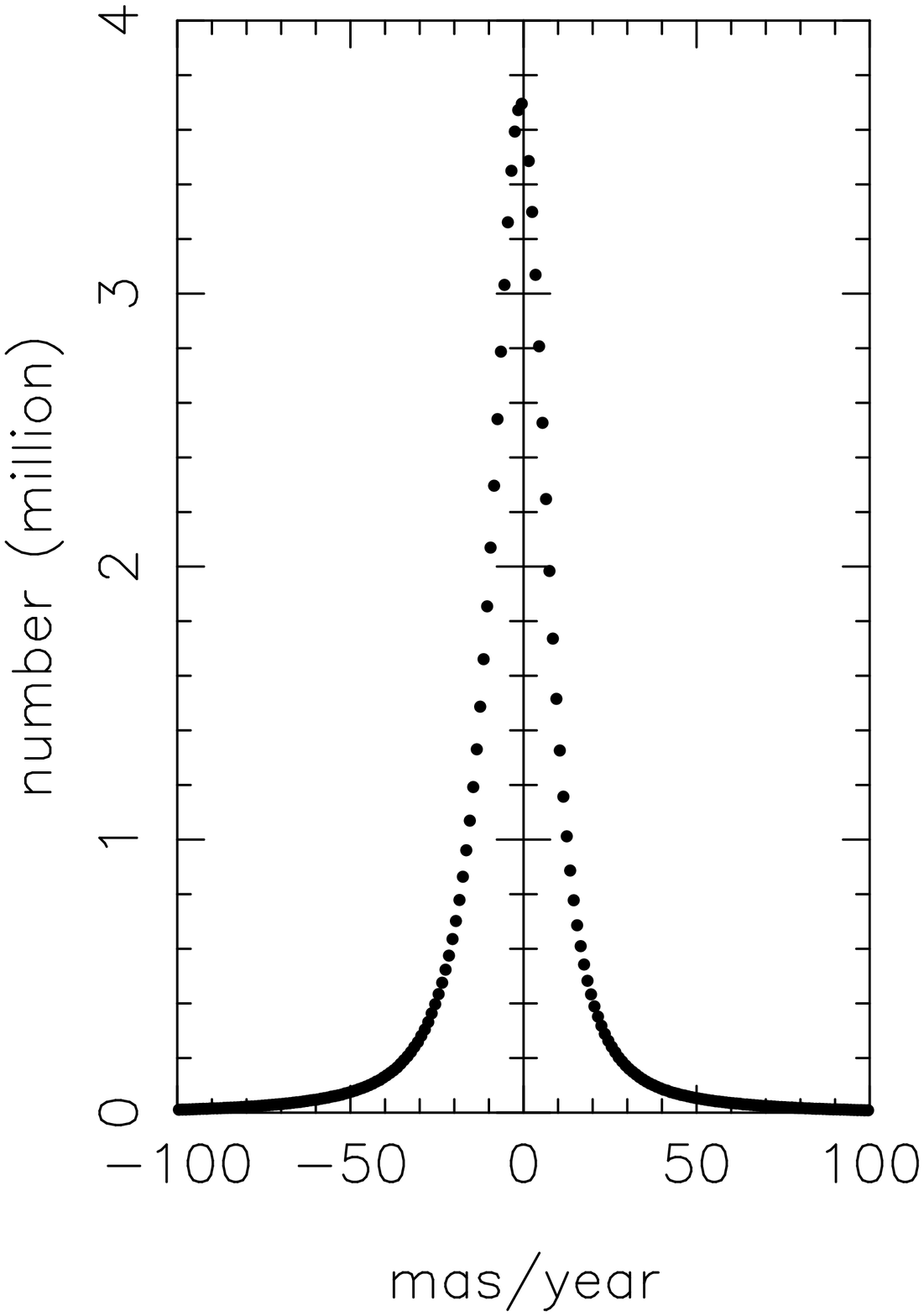}
\caption{Distribution of the UCAC3 proper motions for
         the RA (left) and Dec (right) component.}
\end{figure}

\begin{figure}
\epsscale{1.00}
\plotone{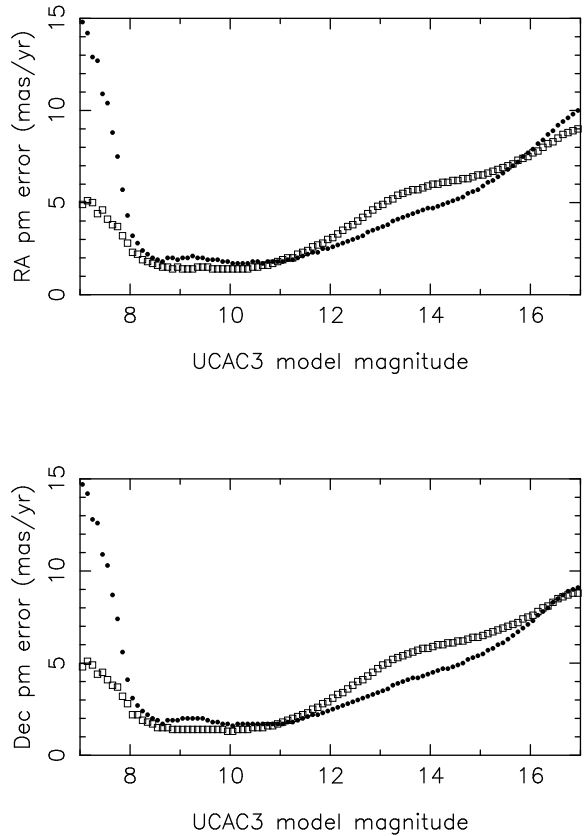}
\caption{Formal, standard errors of UCAC3 proper motions as 
  a function of magnitude, separately for each coordinate.
  The filled dots are for stars in the declination range 
  of $-90^{\circ}$ to $-20^{\circ}$, dominated by the
  SPM first epoch data.  The open squares are for the
  rest of the sky dominated by the SuperCOSMOS first epoch
  data.}
\end{figure}

\begin{figure}
\epsscale{1.00}
\plottwo{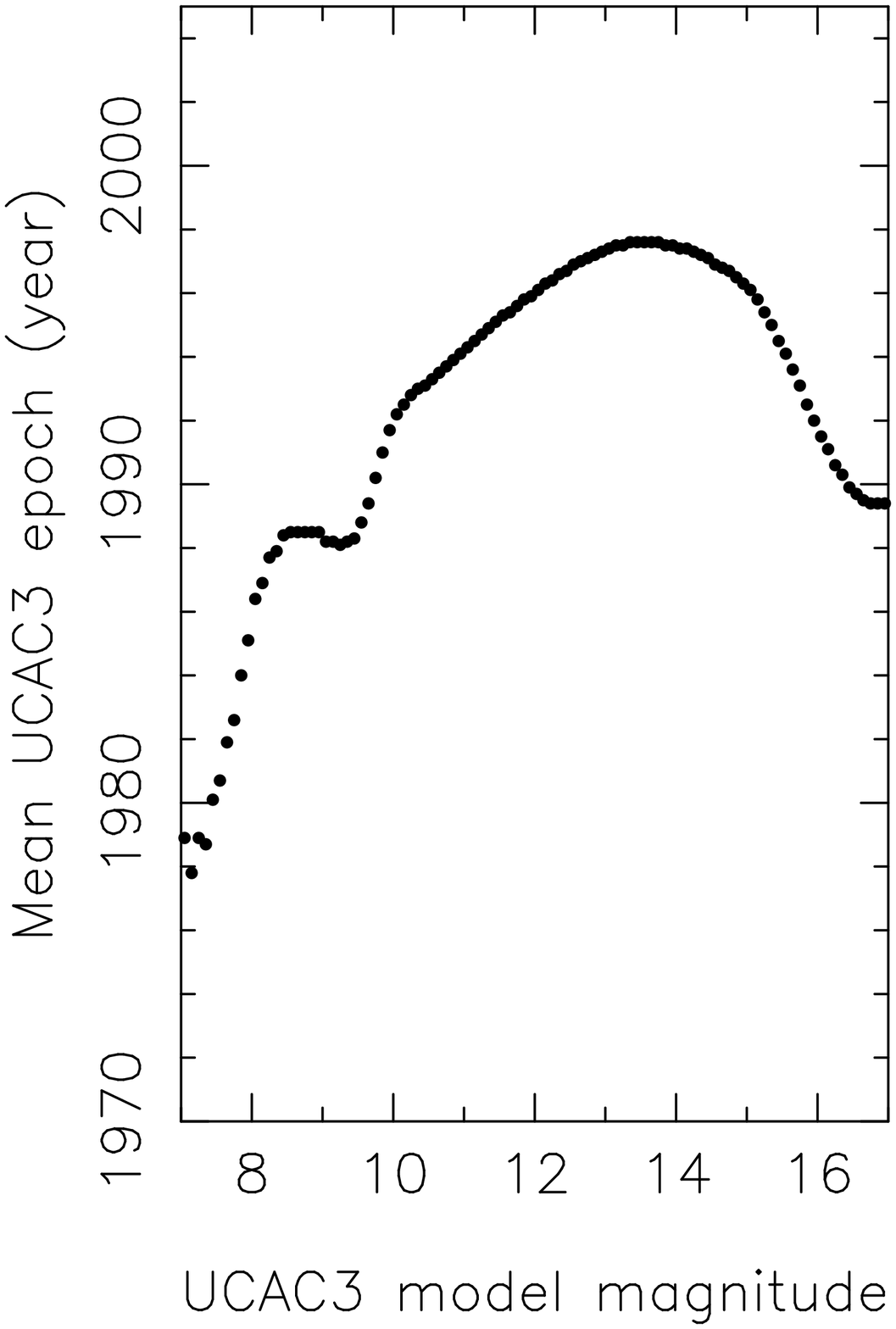}{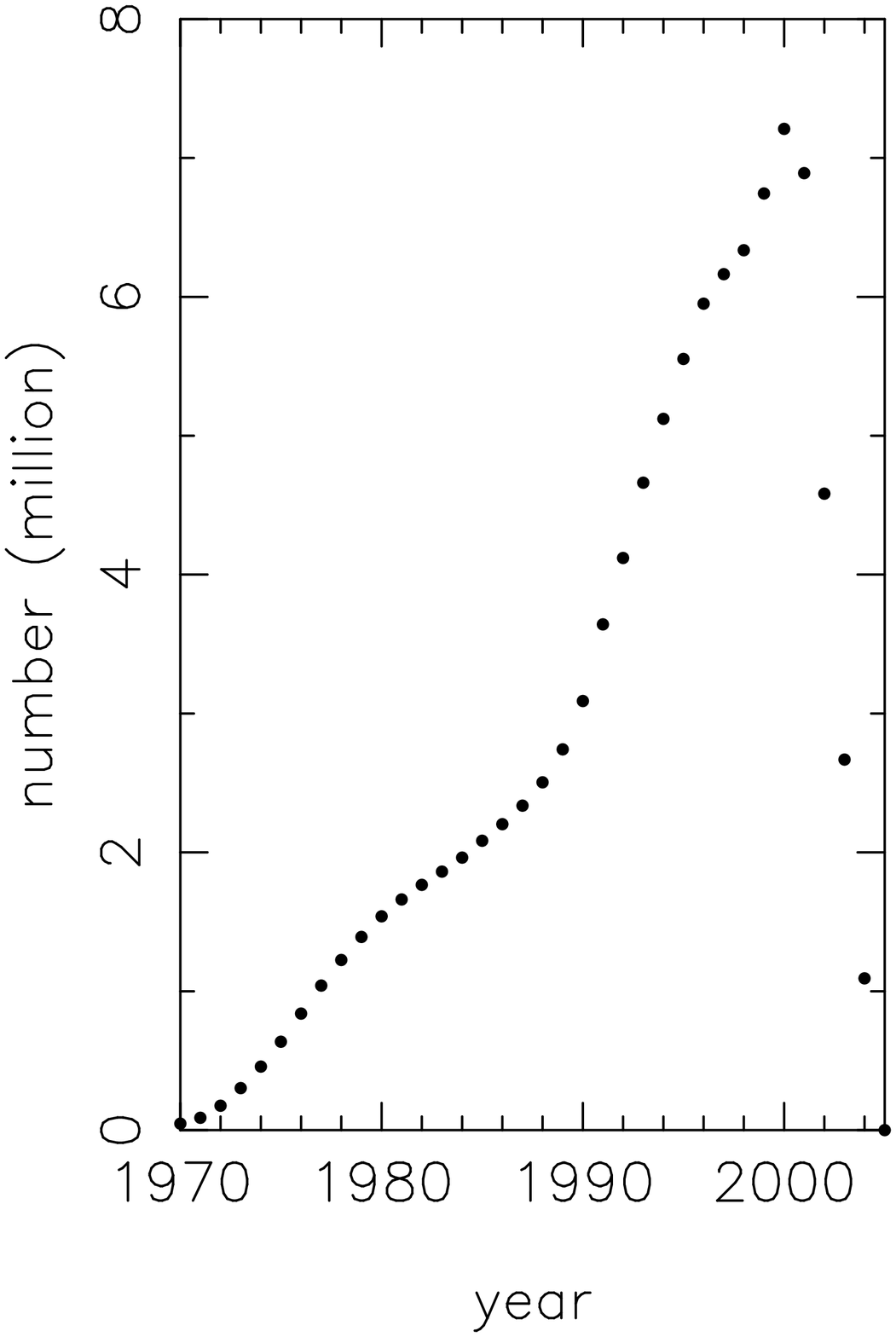}
\caption{Distribution of the mean epoch of all UCAC3 stars 
  as a function of magnitude (left) and as histogram (right).}
\end{figure}


\begin{figure}
\epsscale{.80}
\plotone{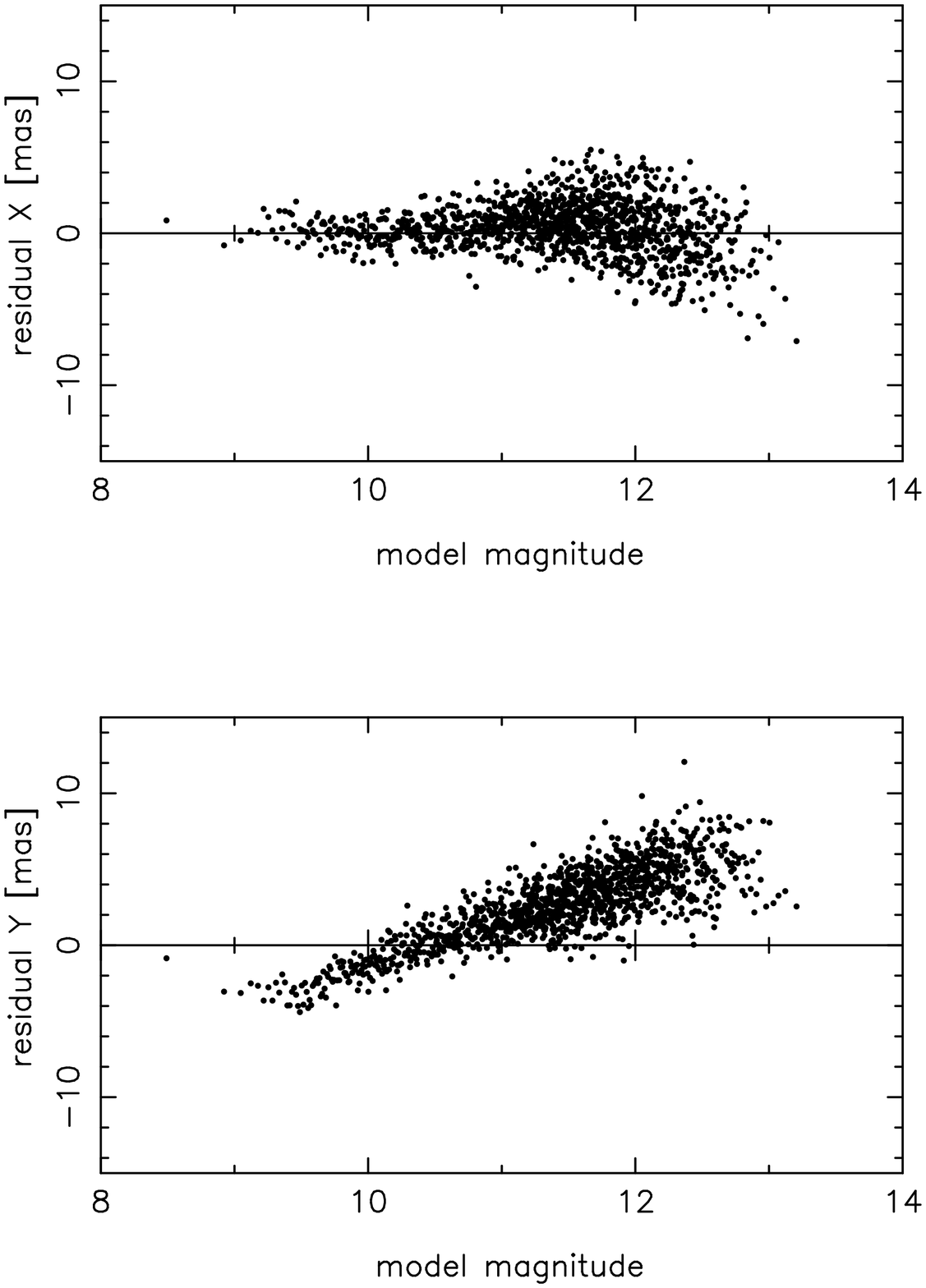}
\caption{Residuals of UCAC frames taken at CTIO with the telescope
 on the West side of the pier, reduced with Tycho-2 reference stars,
 final reductions. Each dot represents the mean over 5000 residuals.} 
\end{figure}

\begin{figure}
\epsscale{.80}
\plotone{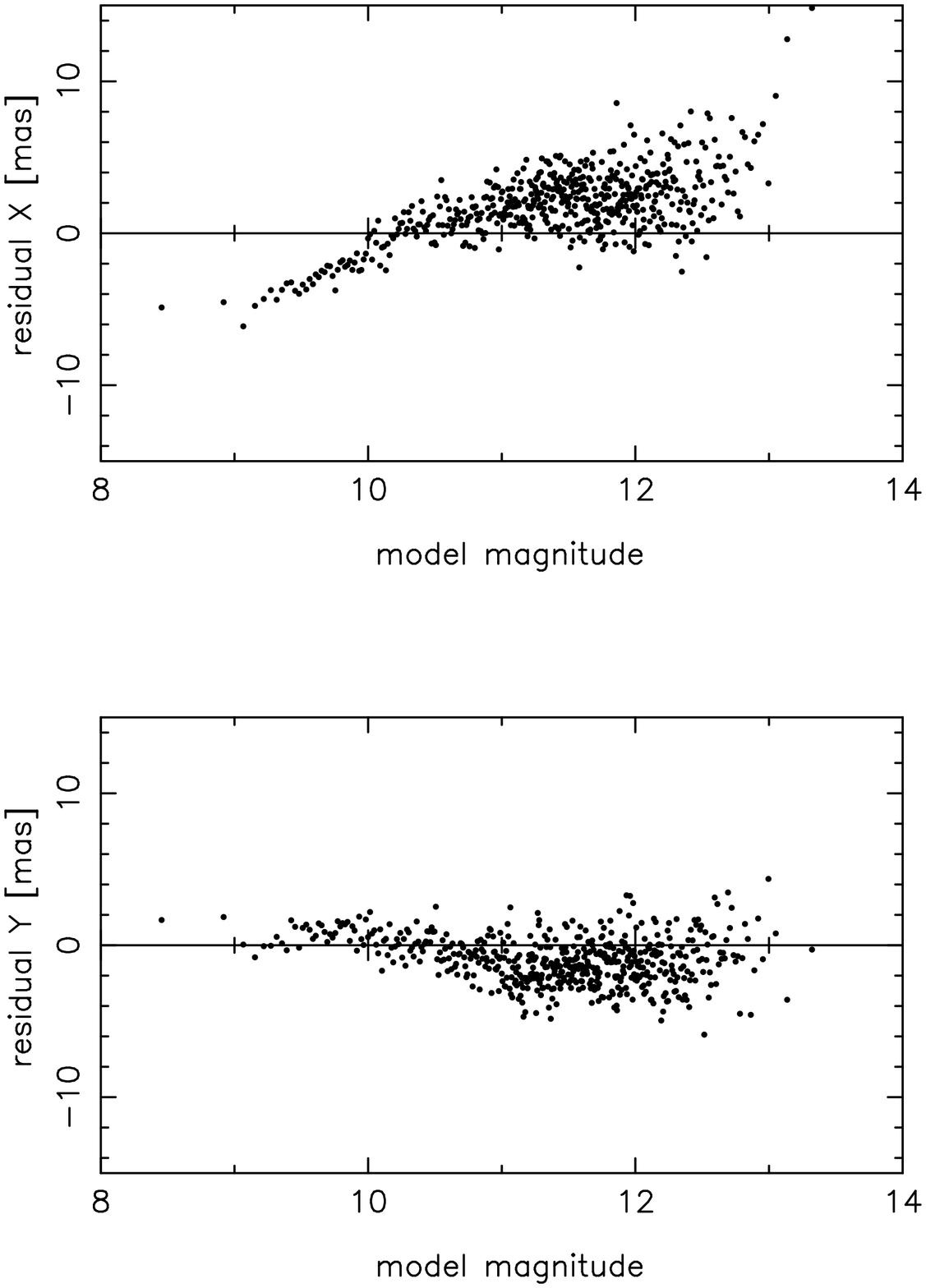}
\caption{Residuals of UCAC frames taken at NOFS with the telescope
 on the East side of the pier, reduced with Tycho-2 reference stars,
 final reductions. Each dot represents the mean over 5000 residuals.} 
\end{figure}


\begin{figure}
\includegraphics[angle=0,scale=.70]{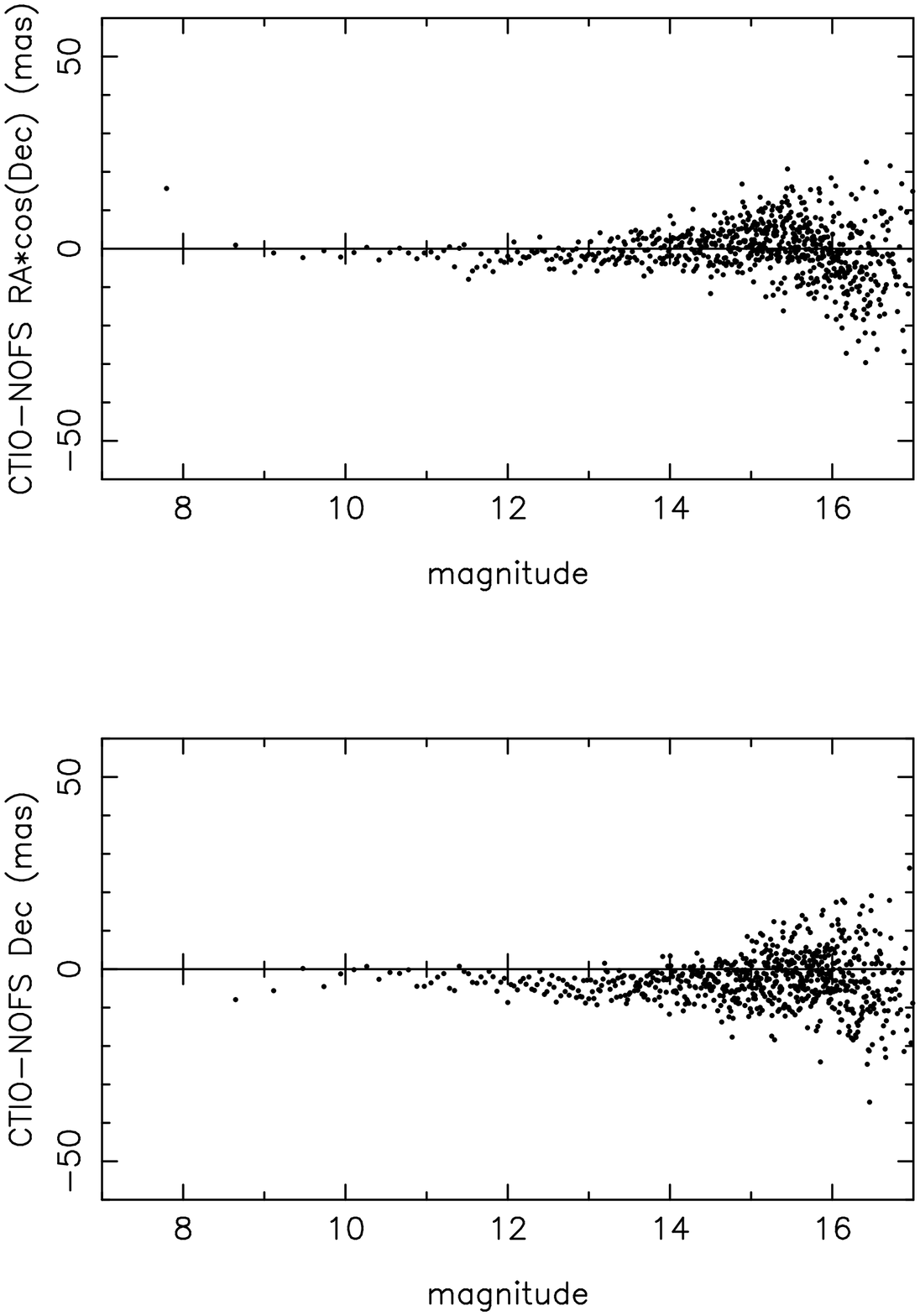}
\caption{Position differences of the same area in the sky as
  observed from CTIO and then from NOFS.  Results are based
  on 1410 CCD frames taken within about 3 months.
  The telescope orientation is flipped by 180 deg between the
  2 data sets, with different systematic error corrections
  applied according to the final pipeline processing.
  Each dot represents the mean over 250 stars.} 
\end{figure}

\clearpage

\begin{figure}
\includegraphics[angle=-90,scale=.70]{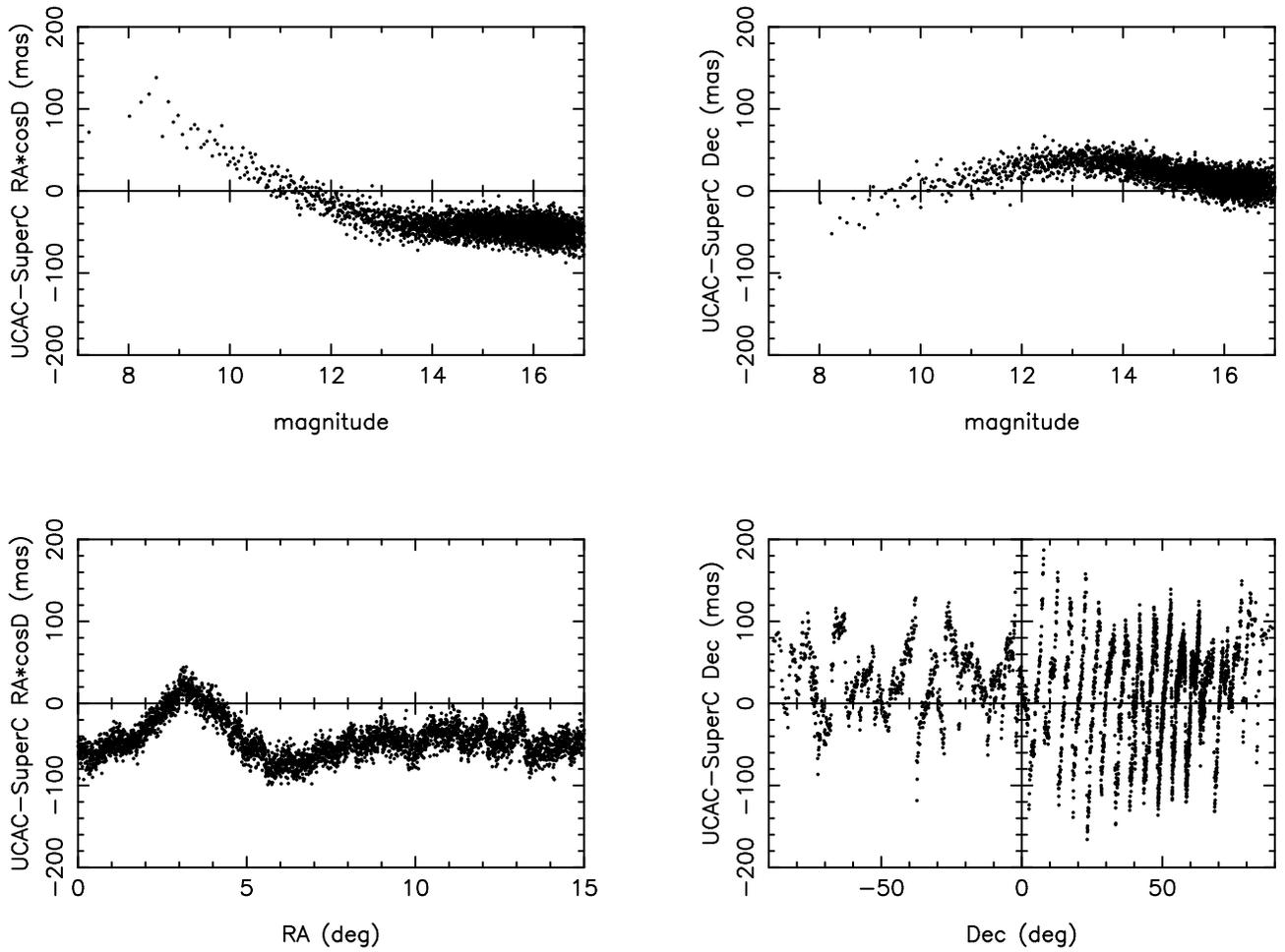}
\caption{Position differences UCAC3 $-$ SuperCOSMOS at epoch of
  UCAC by applying SuperCOSMOS proper motions.
  These data are for a slice along all declinations for RA = 0 to 1 hour
  plotted as a function of UCAC3 model magnitude (top), RA and Dec.
  Each dot represents the mean over 400 stars.} 
\end{figure}

\clearpage

\begin{figure}
\includegraphics[angle=-90,scale=.70]{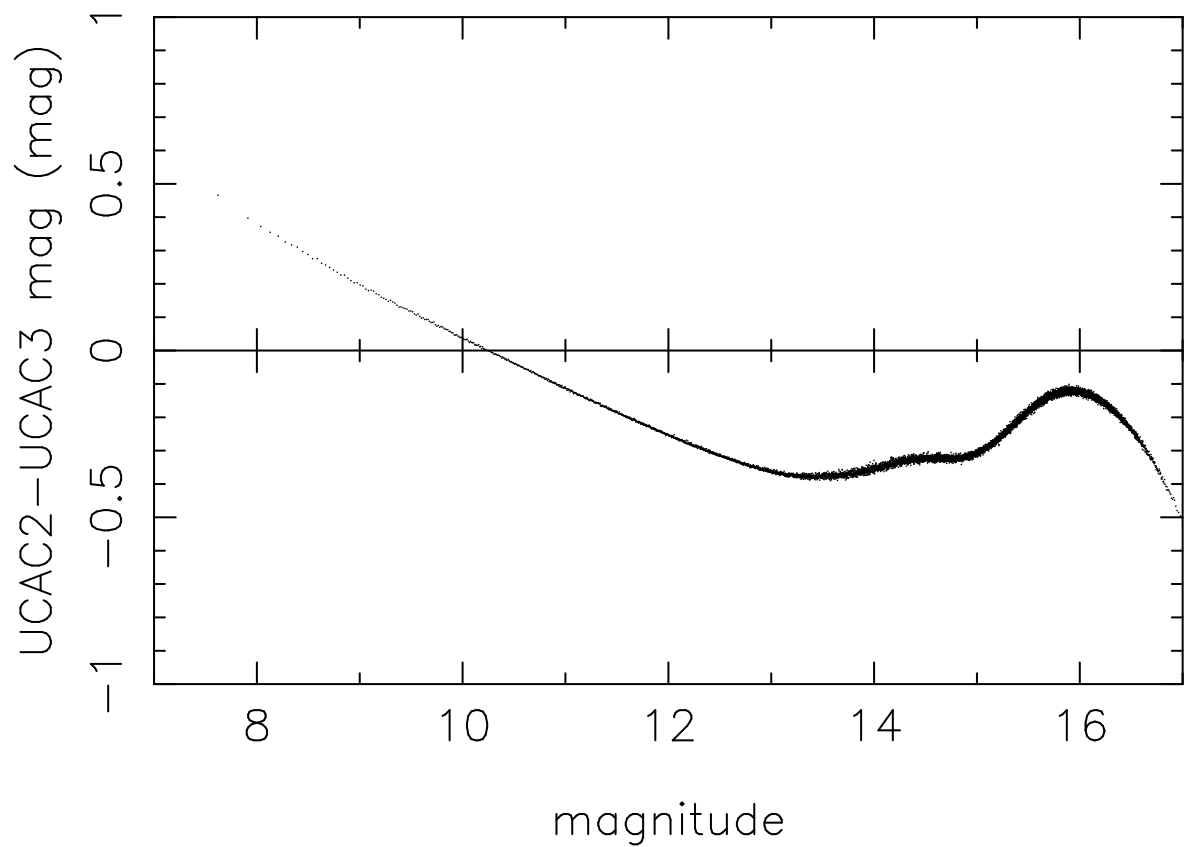}
\caption{Magnitude differences UCAC2 $-$ UCAC3 (model fit) as
  function of UCAC3 magnitude.  The data shown are for the 
  northern hemisphere, the south looks very similar.}
\end{figure}

\clearpage

\begin{figure}
\includegraphics[angle=0,scale=.50]{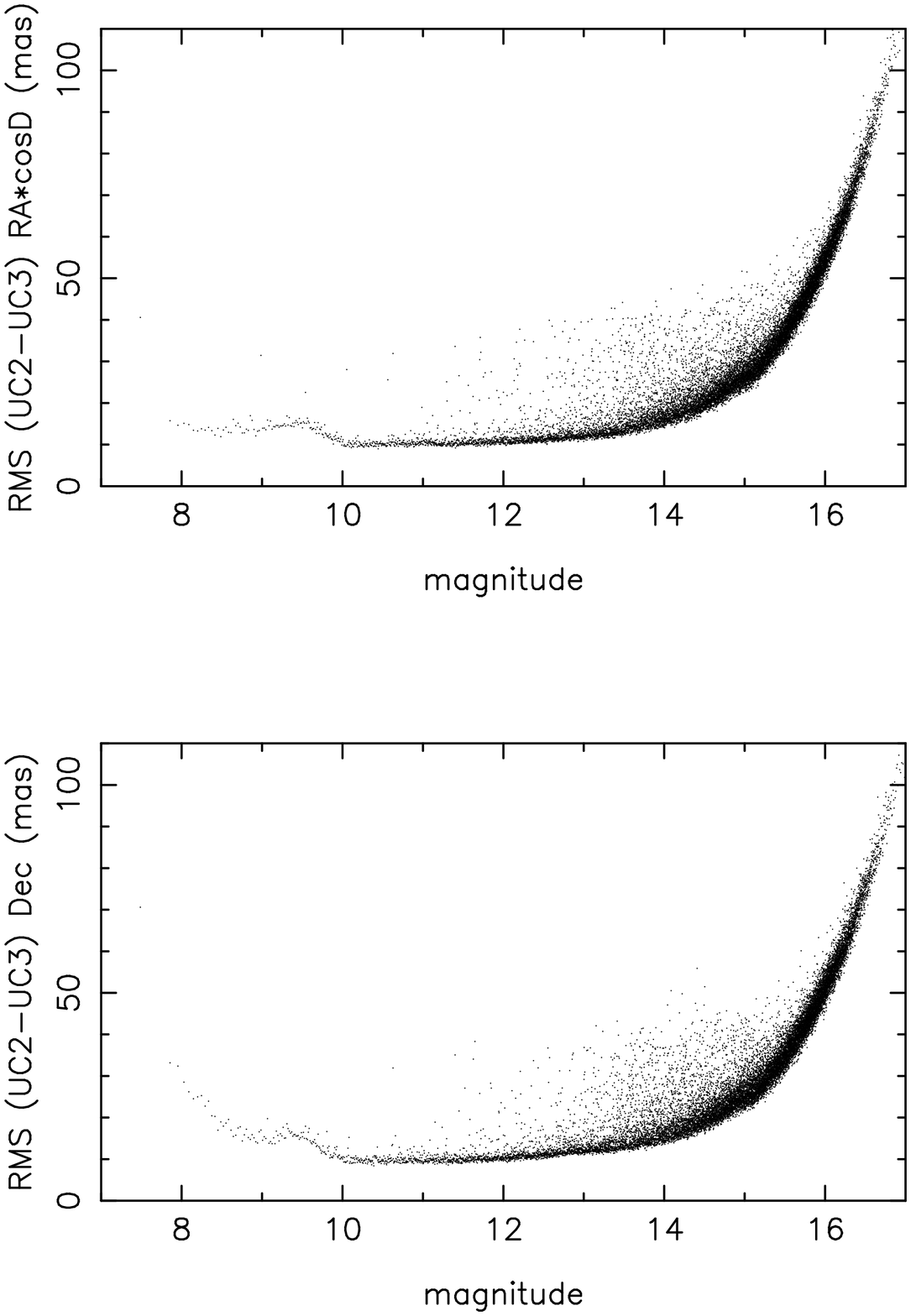}
\caption{RMS position differences UCAC2 $-$ UCAC3 at epoch 2000.0
  for stars on the southern hemisphere.  The upper diagram shows
  results for the RA component and the lower for Dec.}
\end{figure}

\begin{figure}
\includegraphics[angle=0,scale=.50]{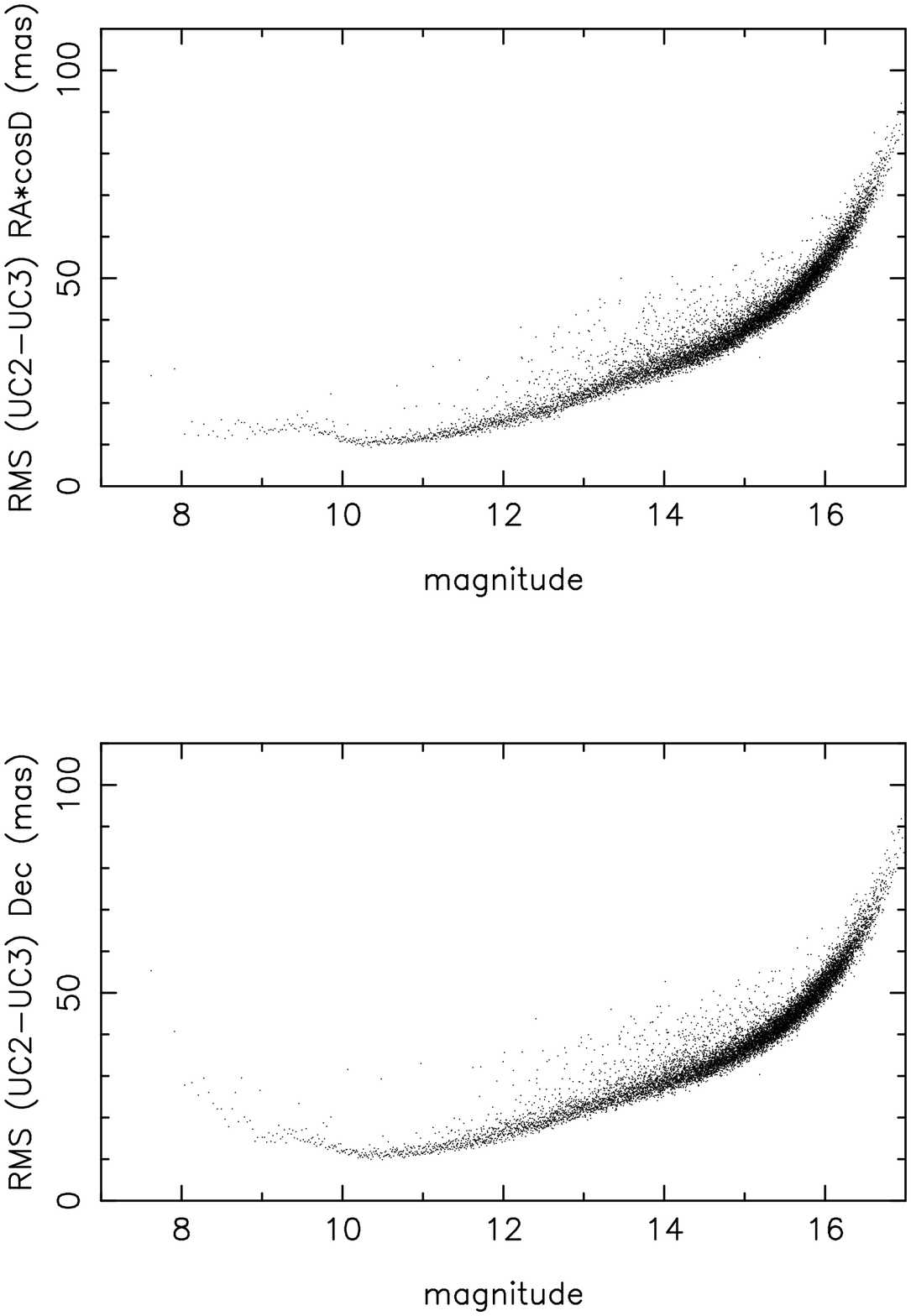}
\caption{Similarly as the previous figure for the northern hemisphere.}
\end{figure}

\clearpage

\begin{figure}
\includegraphics[angle=0,scale=.50]{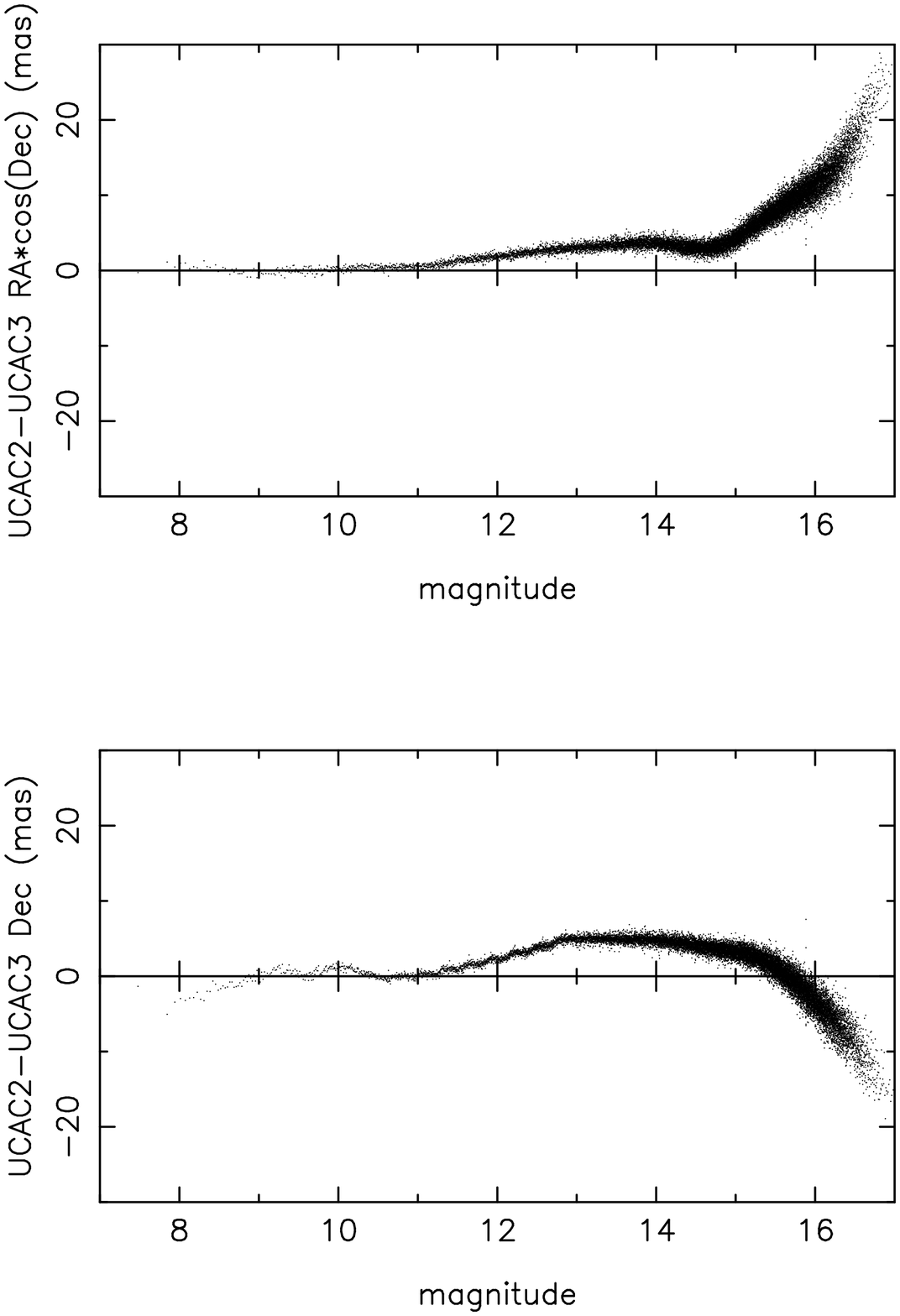}
\caption{Position differences UCAC2 $-$ UCAC3 at epoch 2000.0
  as function of magnitude for stars on the southern hemisphere.}
\end{figure}

\begin{figure}
\includegraphics[angle=0,scale=.50]{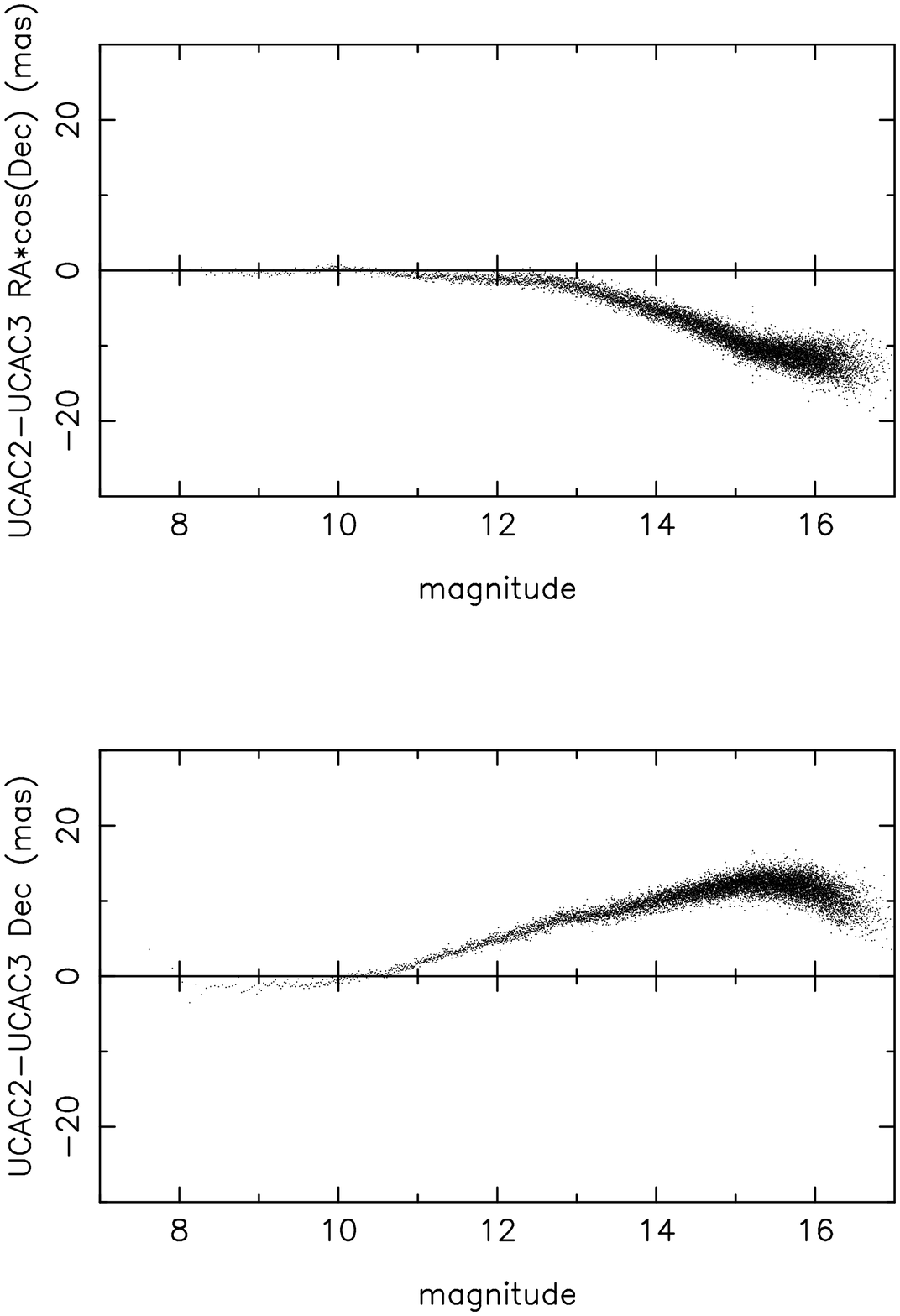}
\caption{Similarly as the previous figure for the northern hemisphere.}
\end{figure}

\clearpage

\begin{figure}
\includegraphics[angle=0,scale=.50]{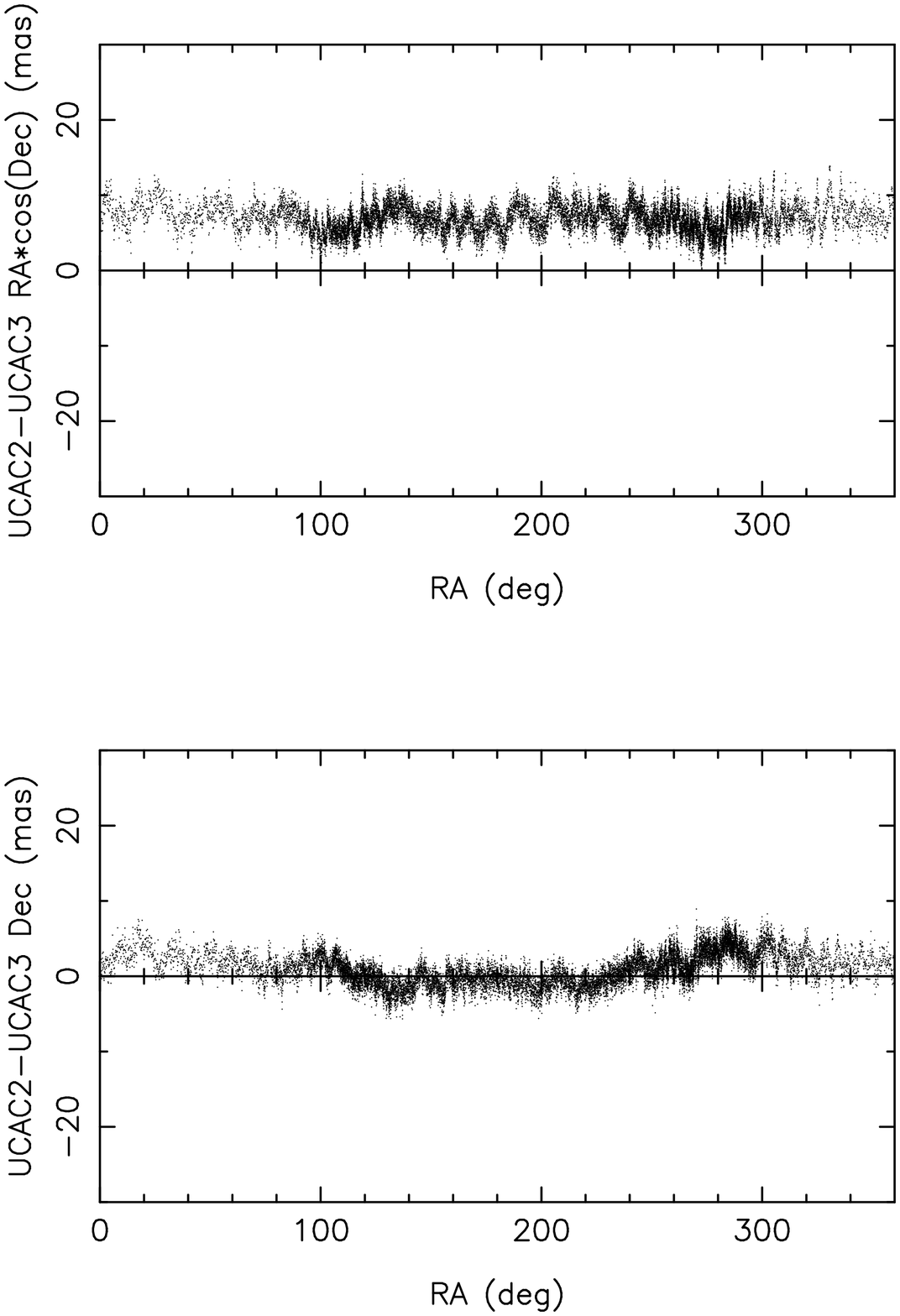}
\caption{Position differences UCAC2 $-$ UCAC3 at epoch 2000.0
  as function of right ascension for stars on the southern hemisphere.}
\end{figure}

\begin{figure}
\includegraphics[angle=0,scale=.50]{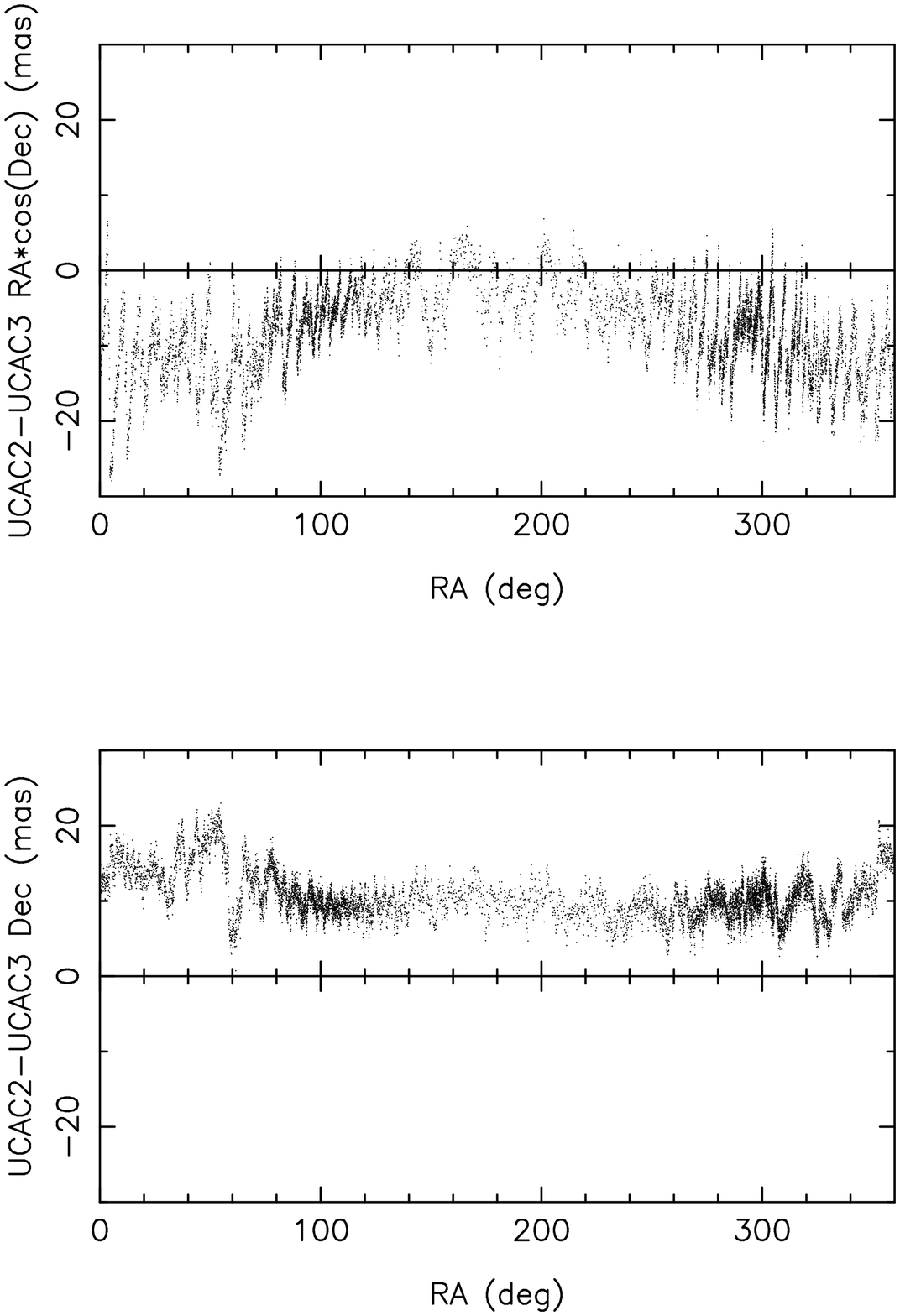}
\caption{Similarly as the previous figure for the northern hemisphere.}
\end{figure}

\clearpage

\begin{figure}
\includegraphics[angle=0,scale=.50]{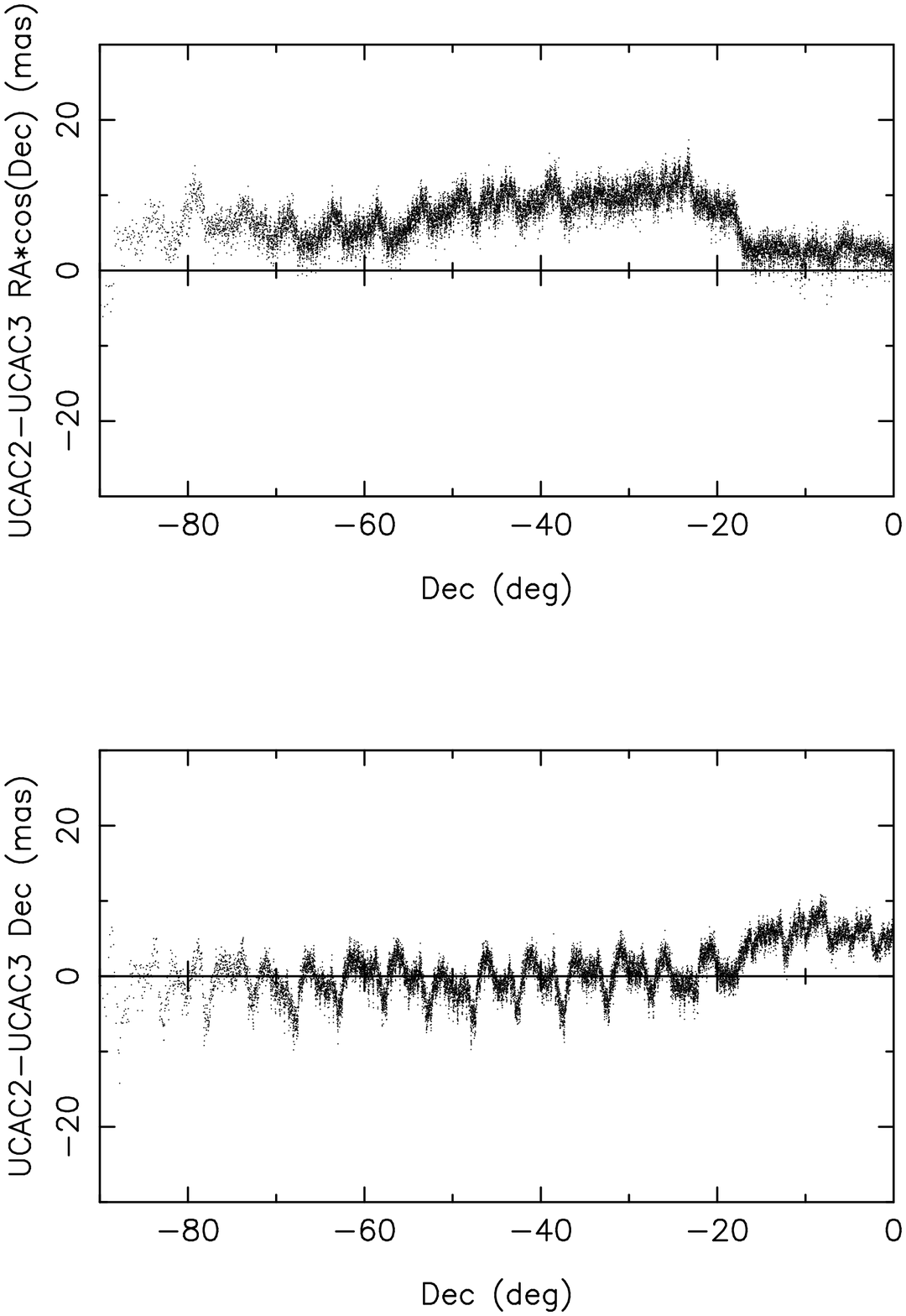}
\caption{Position differences UCAC2 $-$ UCAC3 at epoch 2000.0
  as function of declination for stars on the southern hemisphere.}
\end{figure}

\begin{figure}
\includegraphics[angle=0,scale=.50]{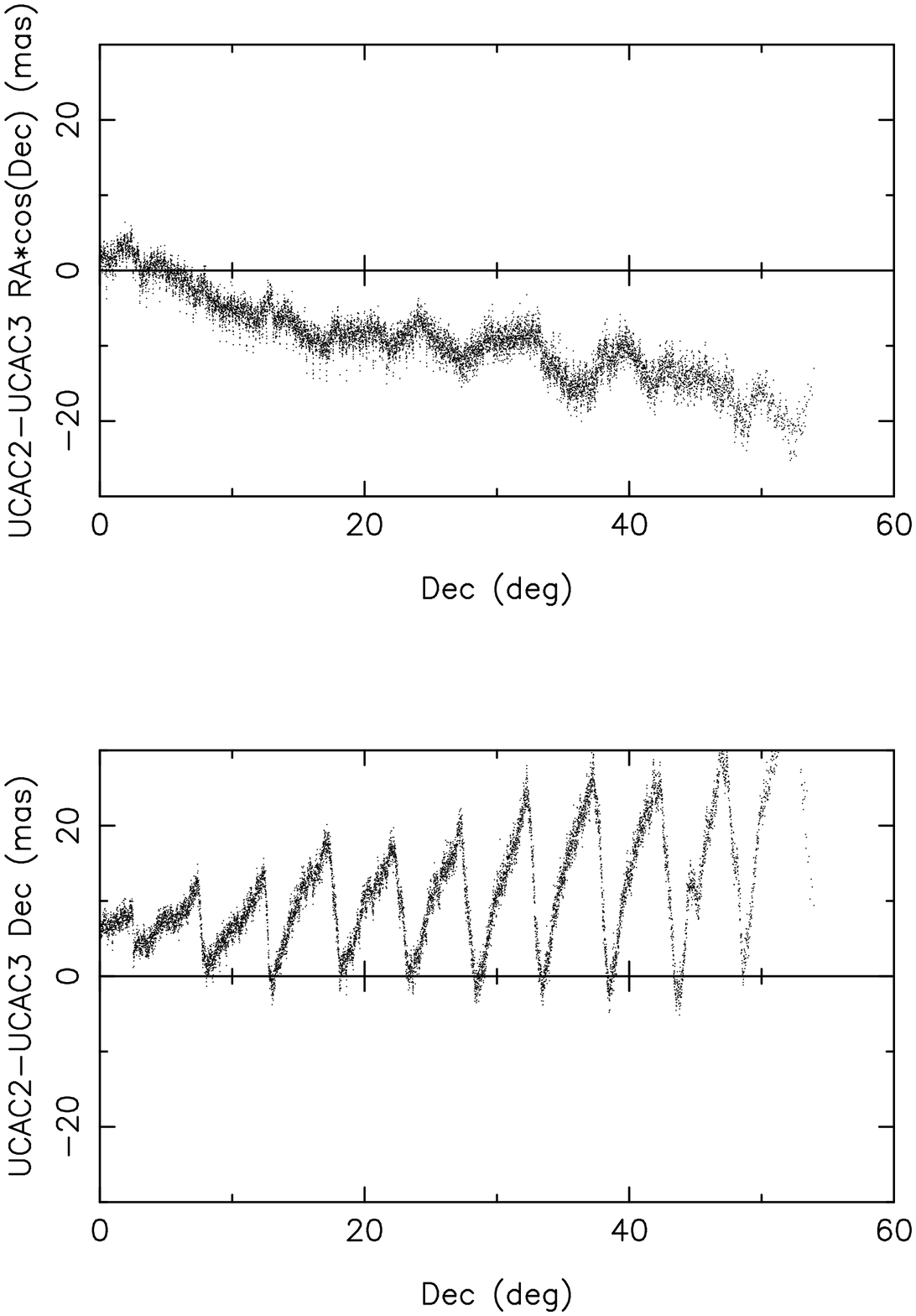}
\caption{Similarly as the previous figure for the northern hemisphere.}
\end{figure}

\clearpage

\begin{figure}
\includegraphics[angle=0,scale=.50]{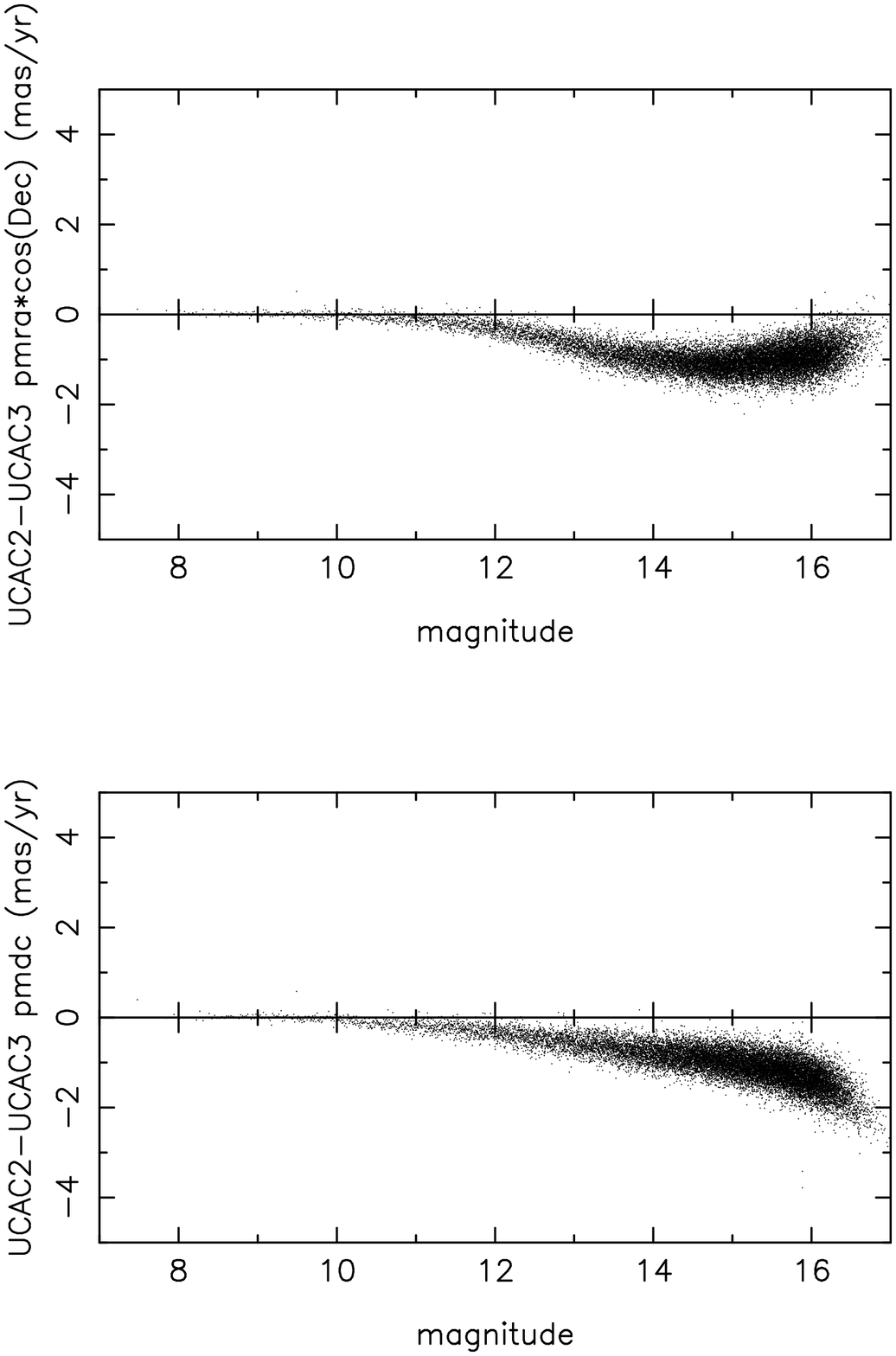}
\caption{Proper motion differences UCAC2 $-$ UCAC3 
  as function of magnitude for stars on the southern hemisphere.}
\end{figure}

\begin{figure}
\includegraphics[angle=0,scale=.50]{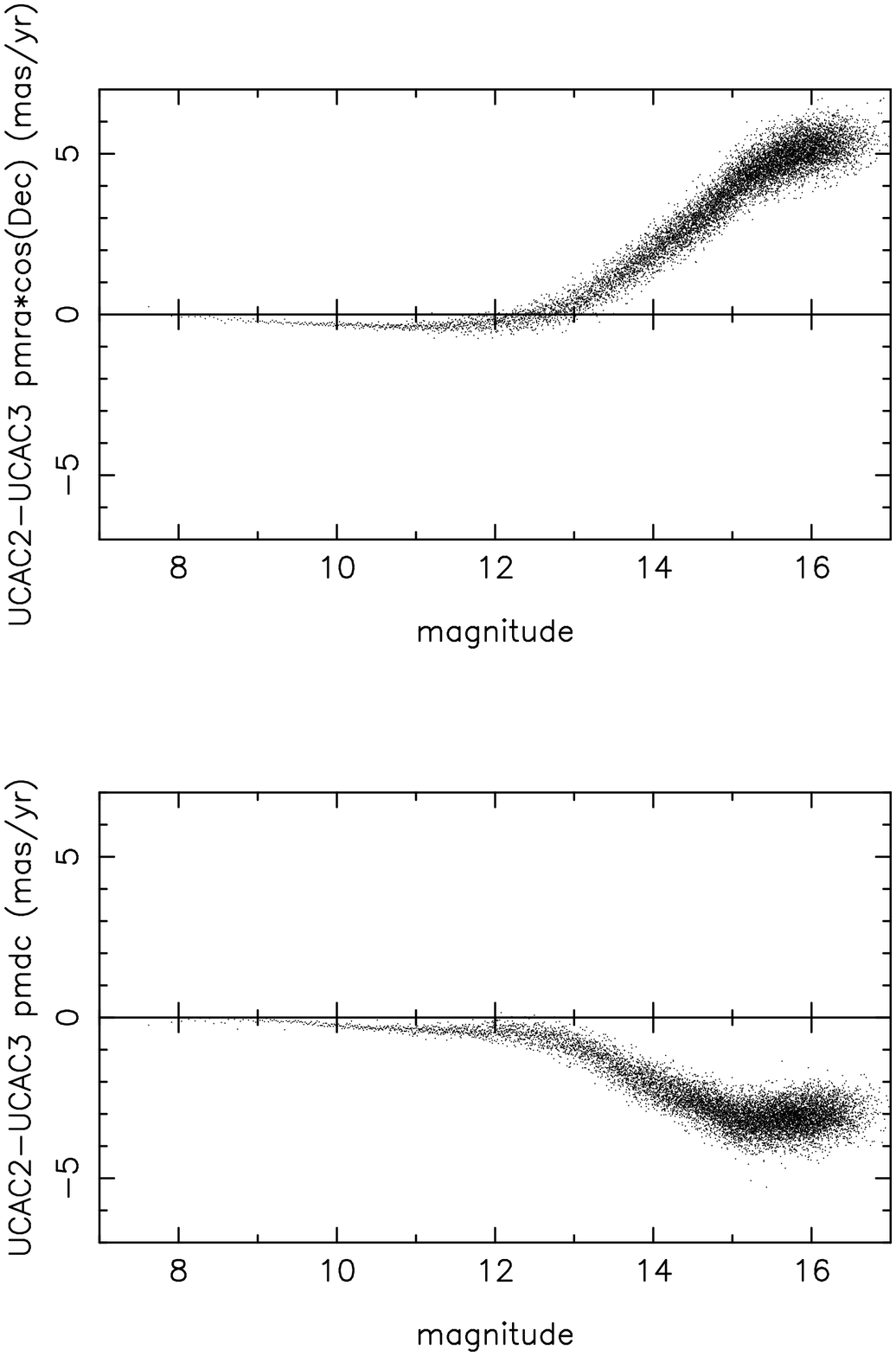}
\caption{Similarly as the previous figure for the northern hemisphere.}
\end{figure}

\clearpage

\begin{figure}
\includegraphics[scale=.50]{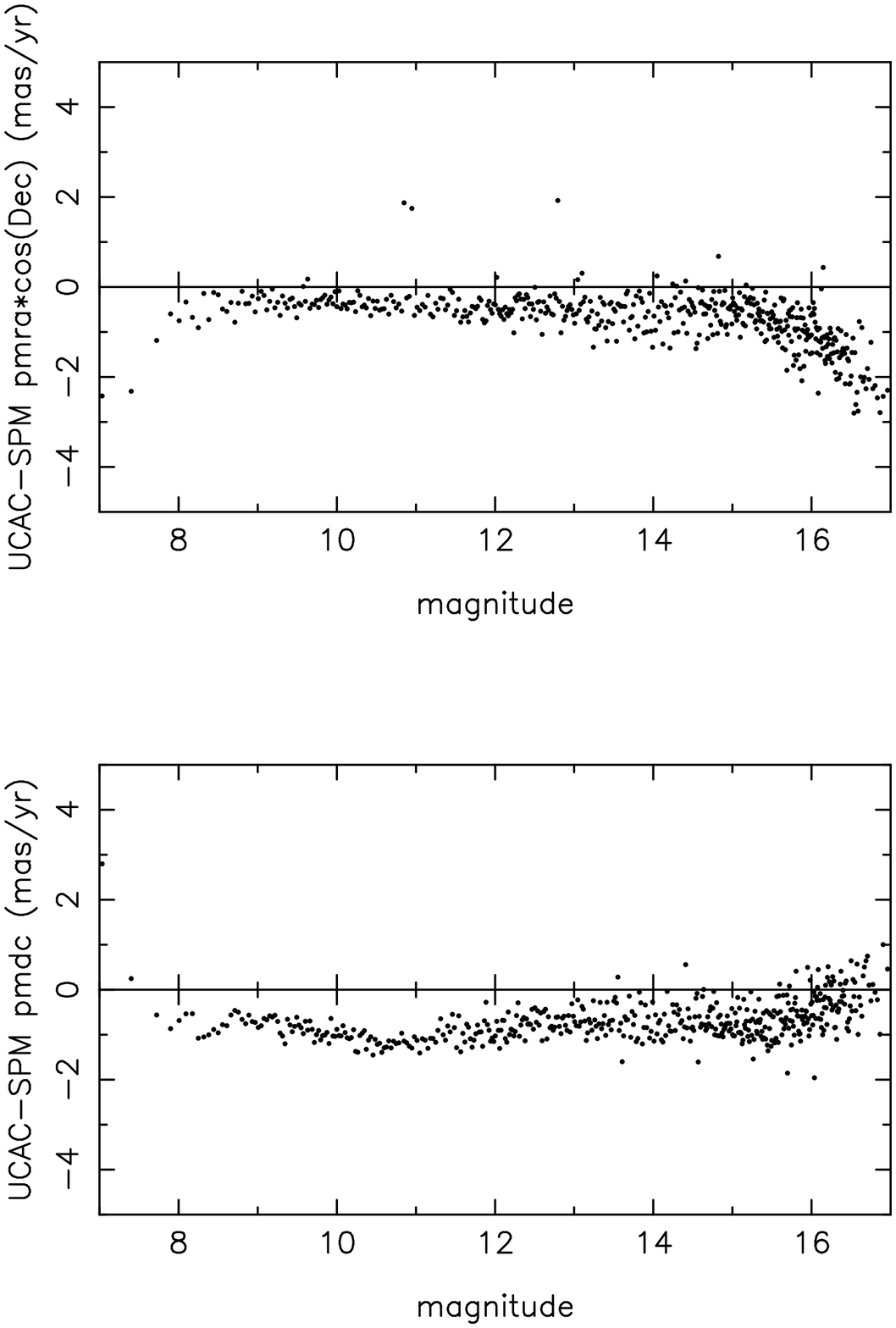}
\caption{Proper motion differences UCAC3 $-$ SPM2.
  For this match of stars the UCAC3 proper motions were used.
  Each dot is the mean over 400 stars.}
\end{figure}

\begin{figure}
\includegraphics[scale=.50]{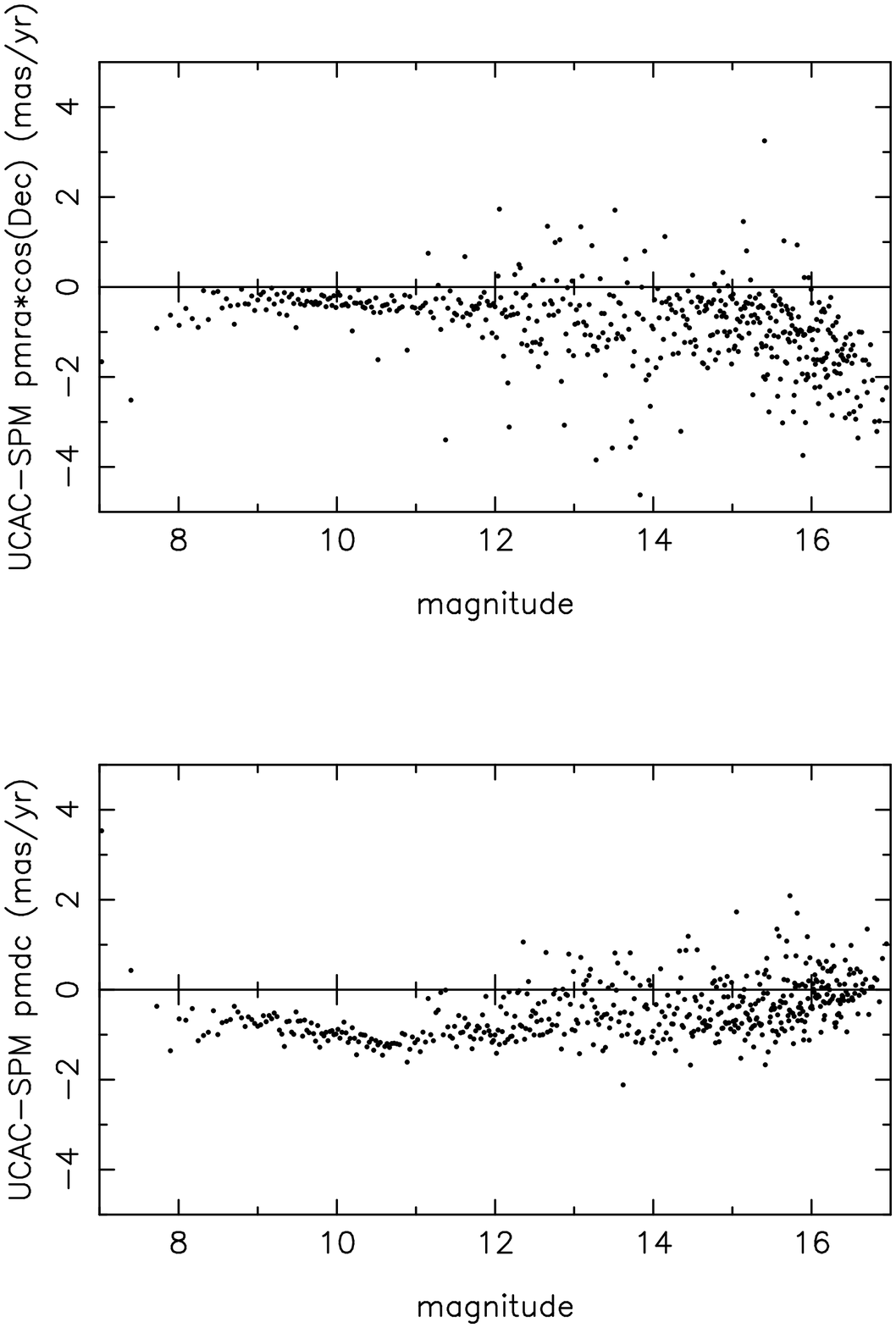}
\caption{Proper motion differences UCAC3 $-$ SPM2.
  For this match of stars the SPM2 proper motions were used.
  Each dot is the mean over 400 stars.}
\end{figure}

\clearpage

\begin{figure}
\includegraphics[scale=.50]{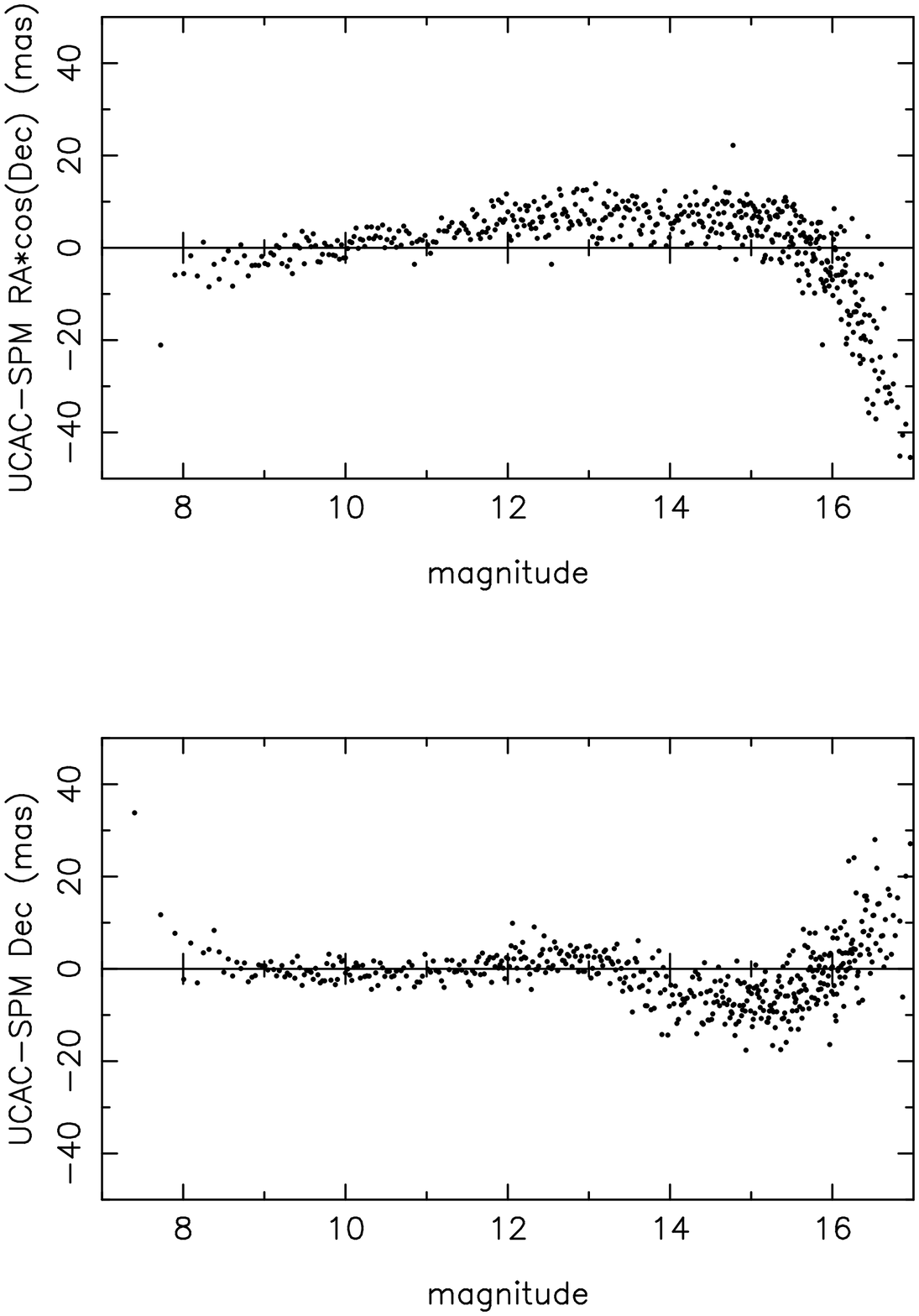}
\caption{Position differences UCAC3 $-$ SPM2 at epoch of
  SPM2 (1991.25) by applying UCAC3 proper motions to the 
  UCAC3 positions (originally at epoch 2000) as function of magnitude.
  Each dot is the mean over 400 stars.}
\end{figure}

\begin{figure}
\includegraphics[scale=.50]{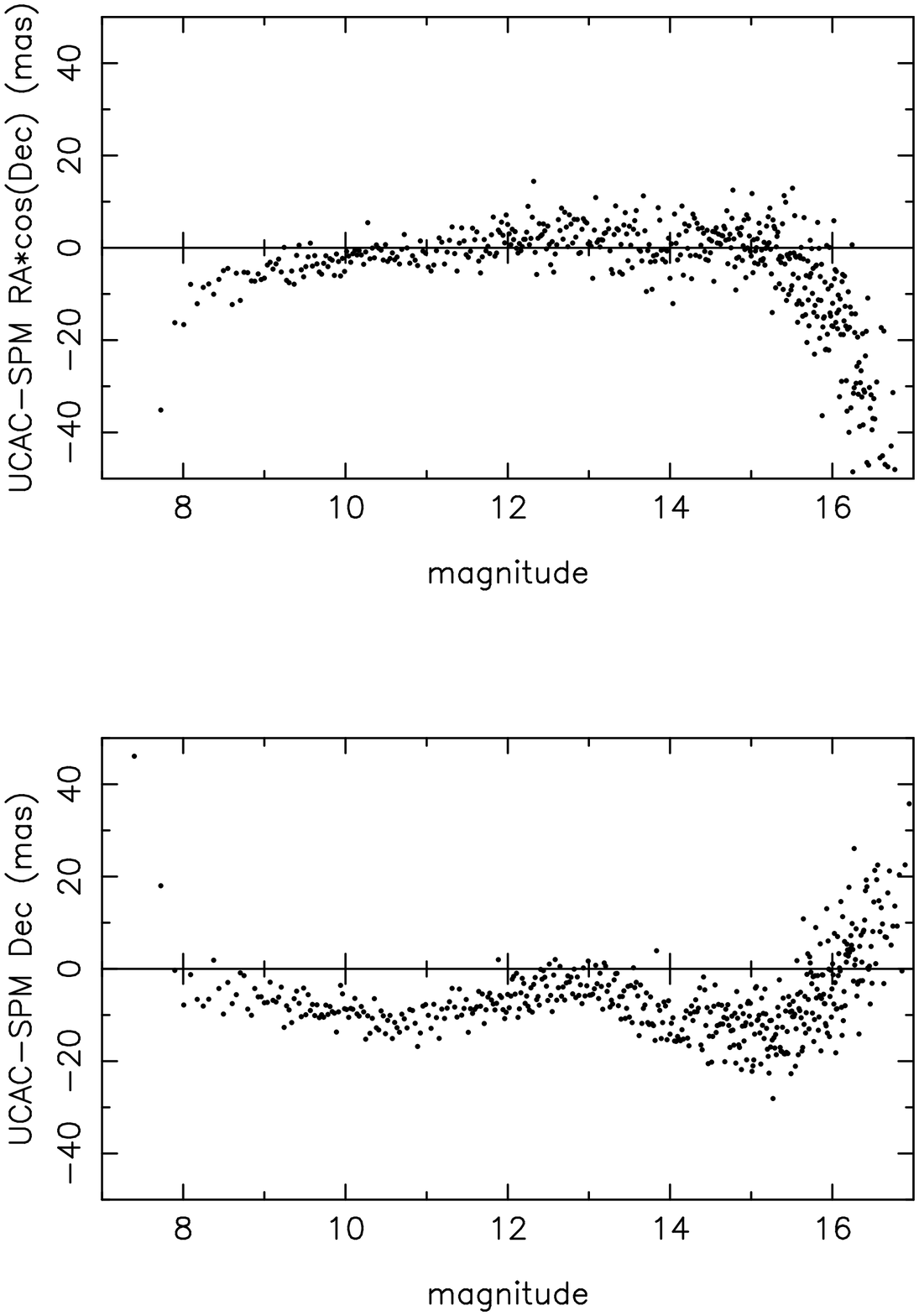}
\caption{Position differences UCAC3 $-$ SPM2 at epoch of
  UCAC by applying SPM2 proper motions, as function of magnitude.
  Each dot is the mean over 400 stars.}
\end{figure}

\clearpage

\begin{figure}
\includegraphics[scale=.50]{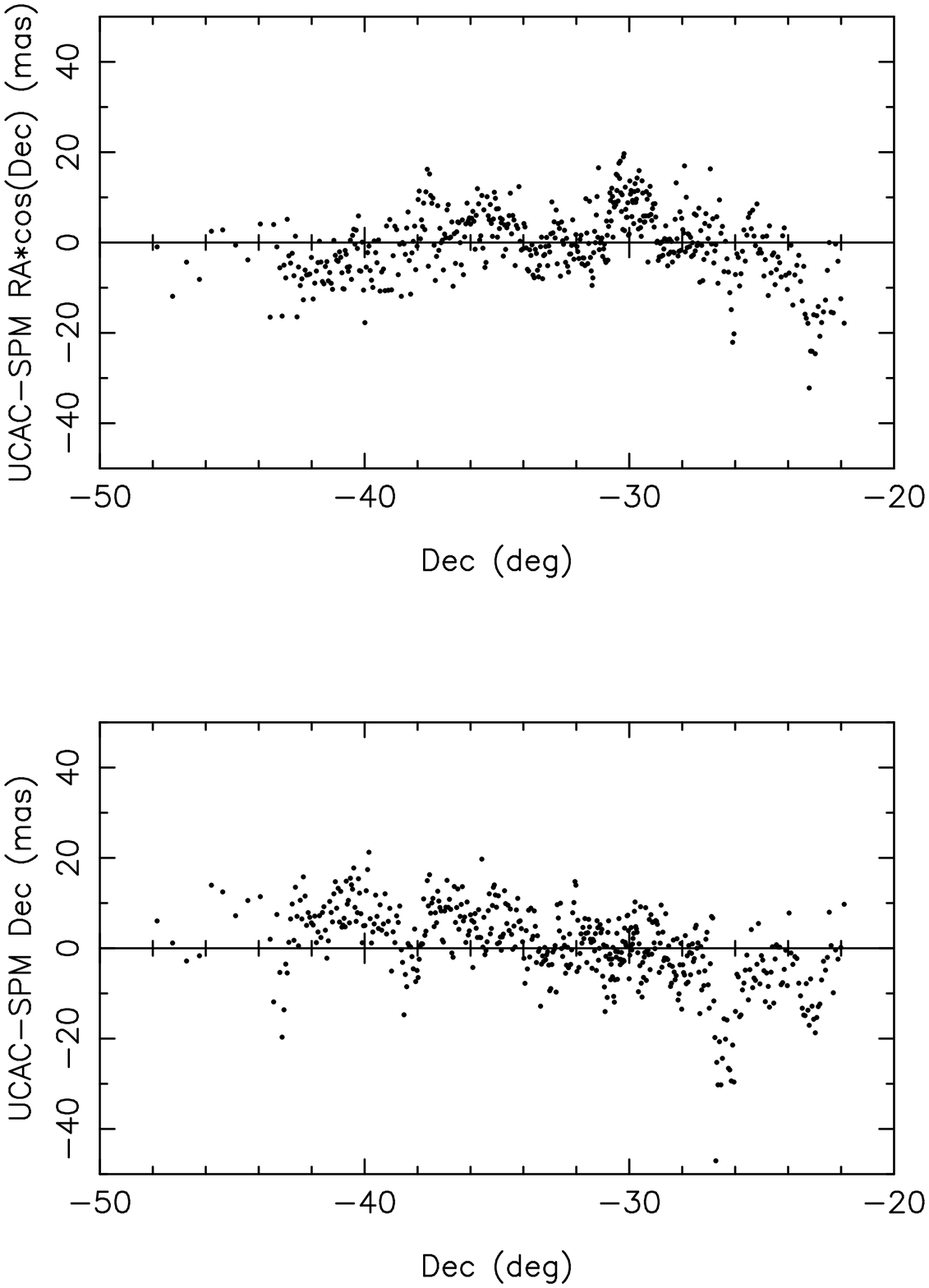}
\caption{Position differences UCAC3 $-$ SPM2 at epoch of
  SPM2 (1991.25) by applying UCAC3 proper motions to the 
  UCAC3 positions (originally at epoch 2000) as function of declination.
  Each dot is the mean over 400 stars.}
\end{figure}

\begin{figure}
\includegraphics[scale=.50]{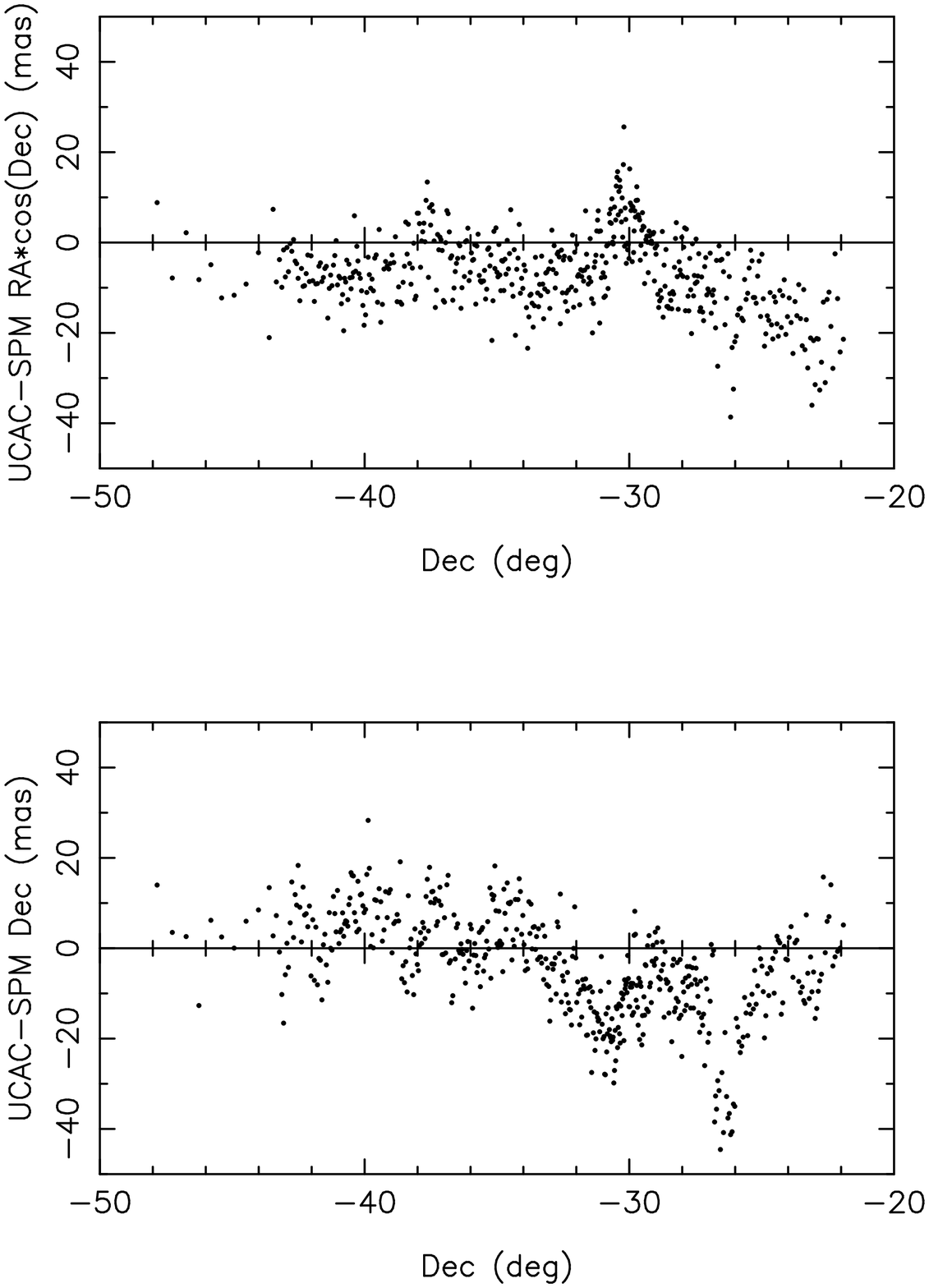}
\caption{Position differences UCAC3 $-$ SPM2 at epoch of
  UCAC by applying SPM2 proper motions, as function of declination.
  Each dot is the mean over 400 stars.}
\end{figure}

\clearpage

\begin{figure}
\includegraphics[scale=.45]{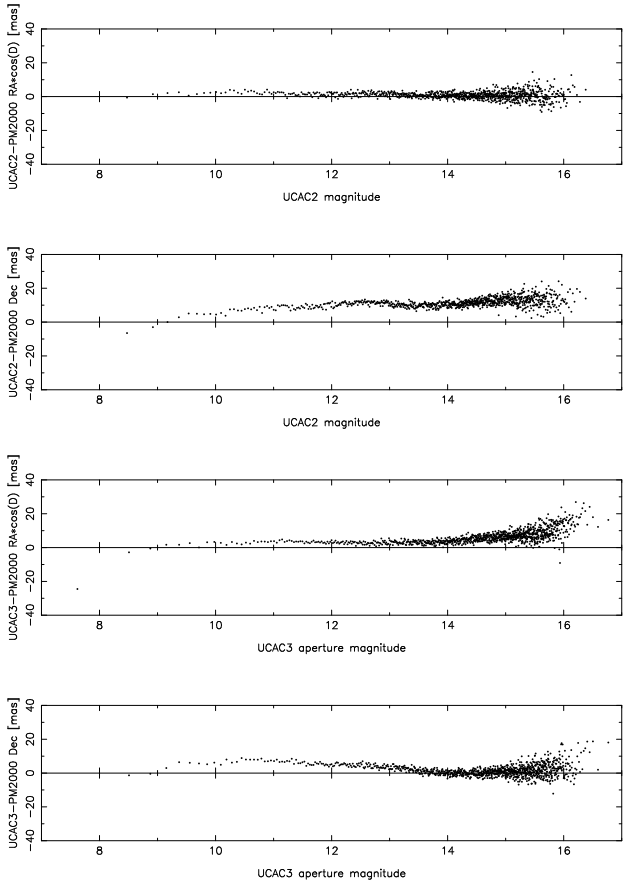}
\caption{Position differences UCAC2 $-$ PM2000 (top) and
  UCAC3 $-$ PM2000 at epoch 2000.0
  using PM2000 proper motions, as function of magnitude.
  These data cover the about $9.5^{\circ}$ to $18.5^{\circ}$ declination
  zone.  Each dot represents the mean over 2500 differences.}
\end{figure}

\begin{figure}
\includegraphics[scale=.50]{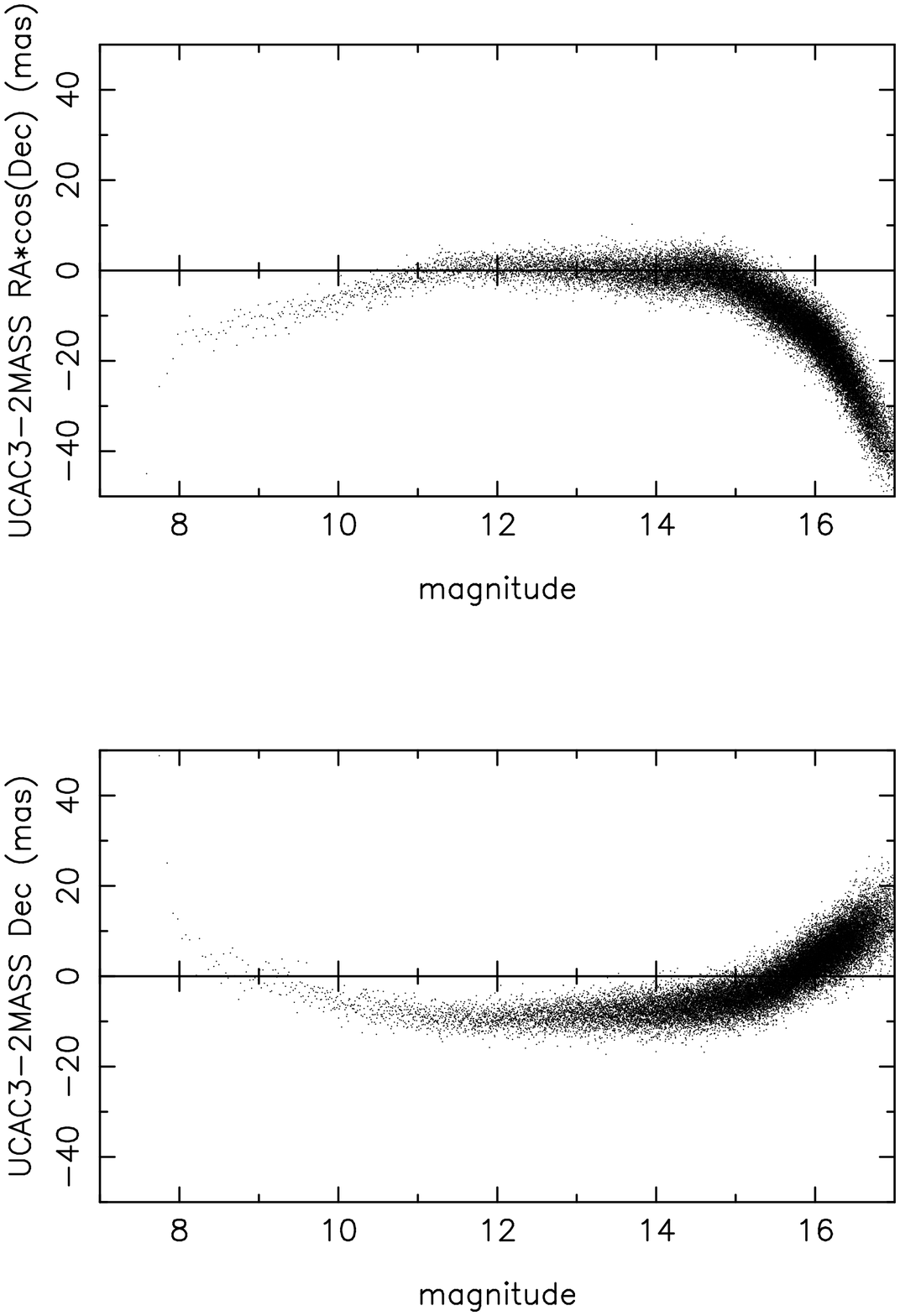}
\caption{Position differences UCAC3 $-$ 2MASS at the 2MASS epoch 
  using UCAC3 proper motions, as function of magnitude.
  These data are for the southern hemisphere.}
\end{figure}

\begin{figure}
\includegraphics[scale=.50]{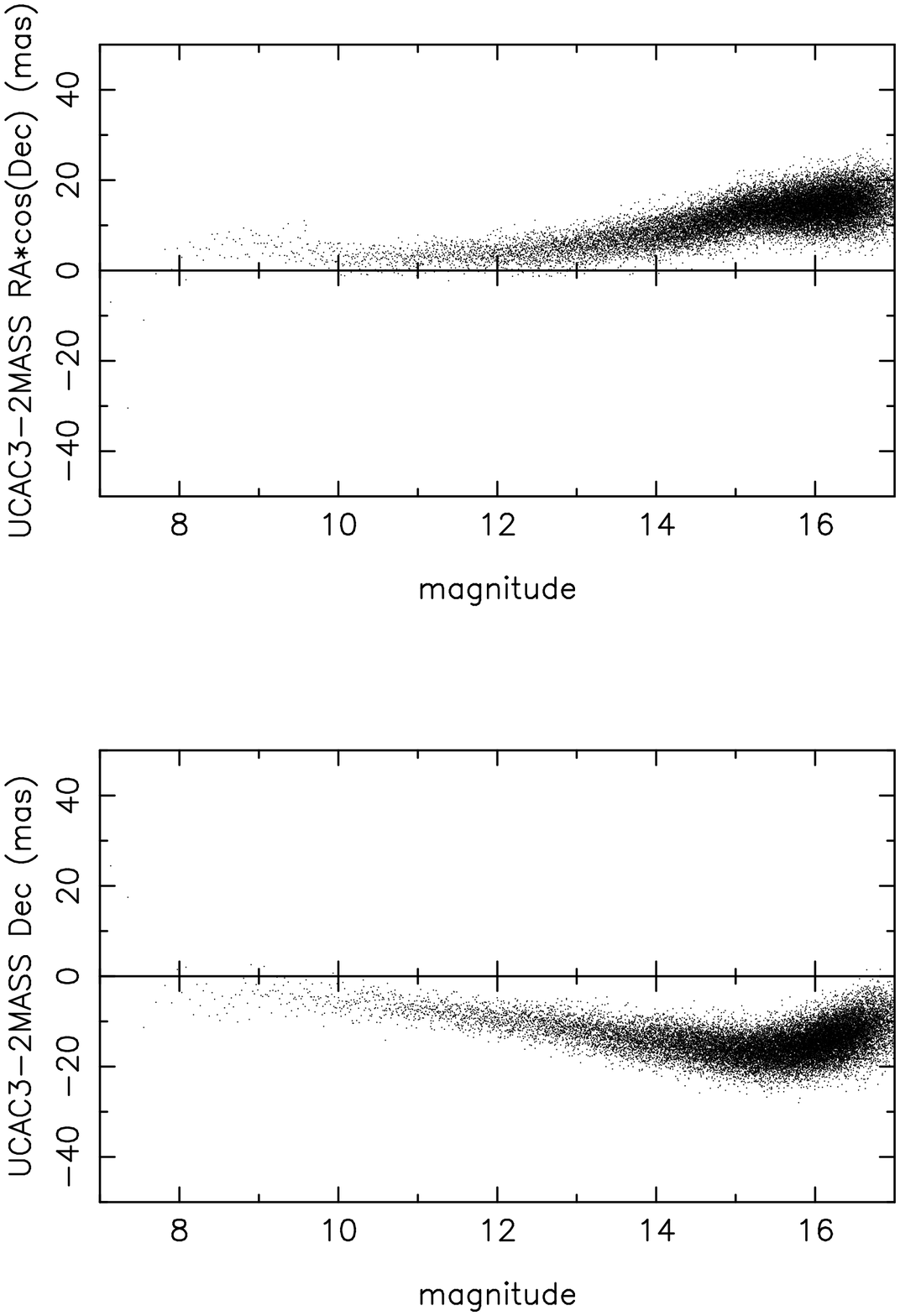}
\caption{Similar to the previous figure but for the northern hemisphere.}
\end{figure}

\clearpage

\begin{figure}
\includegraphics[scale=.50]{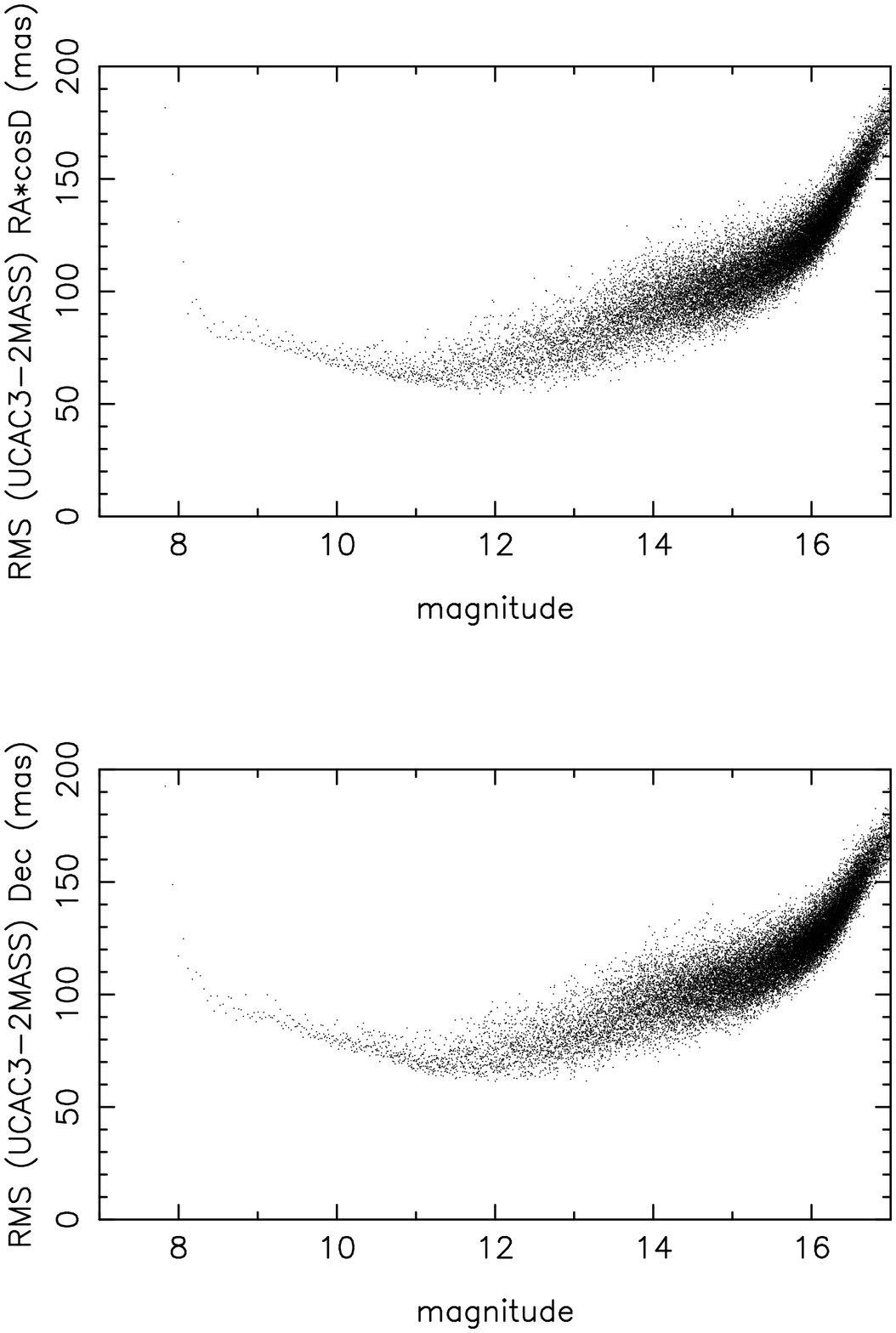}
\caption{RMS position differences UCAC3 $-$ 2MASS at the 2MASS epoch 
  using UCAC3 proper motions, as function of magnitude.
  These data are for the southern hemisphere.}
\end{figure}

\begin{figure}
\includegraphics[scale=.50]{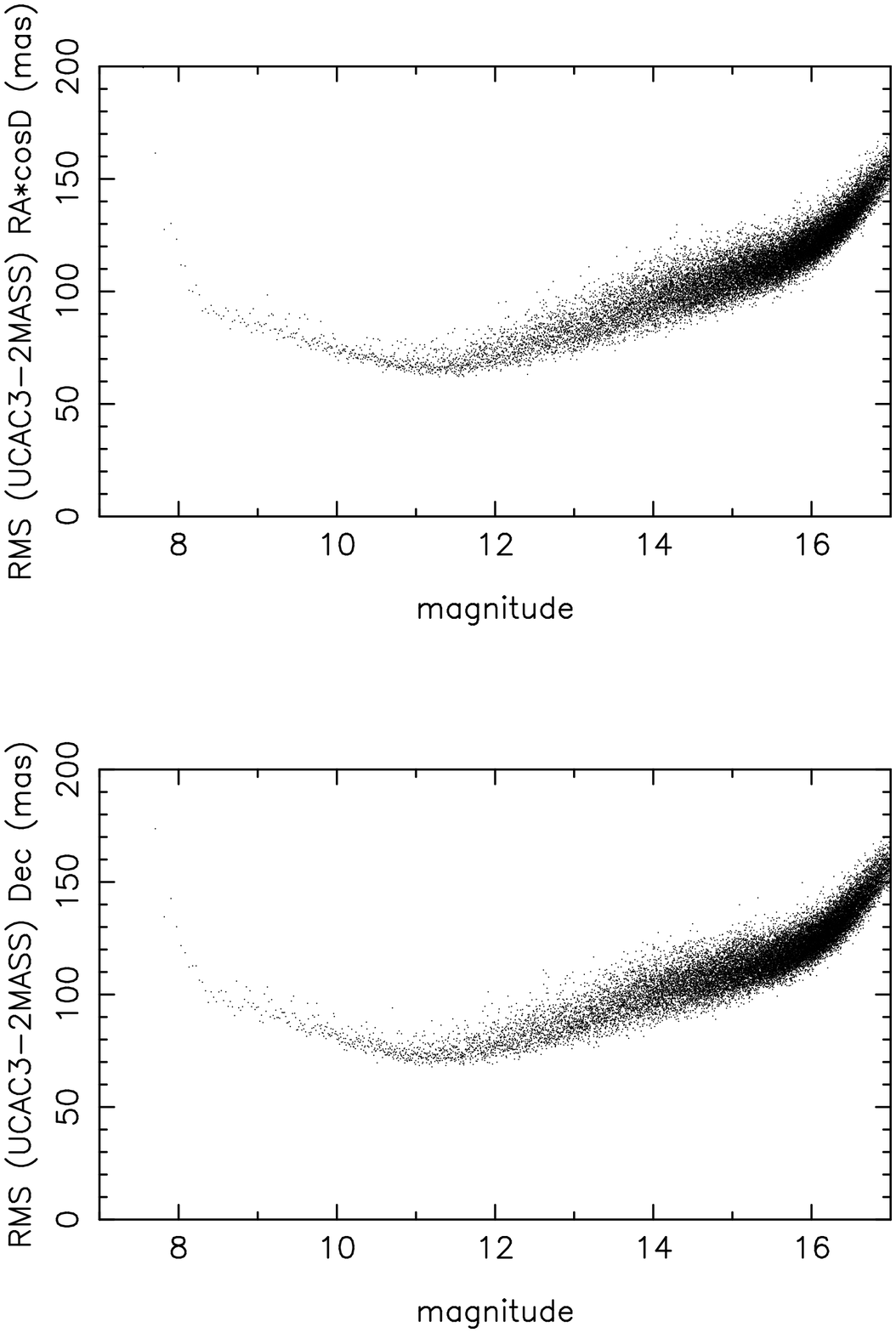}
\caption{Similar to the previous figure but for the northern hemisphere.}
\end{figure}

\clearpage

\begin{figure}
\includegraphics[scale=.50]{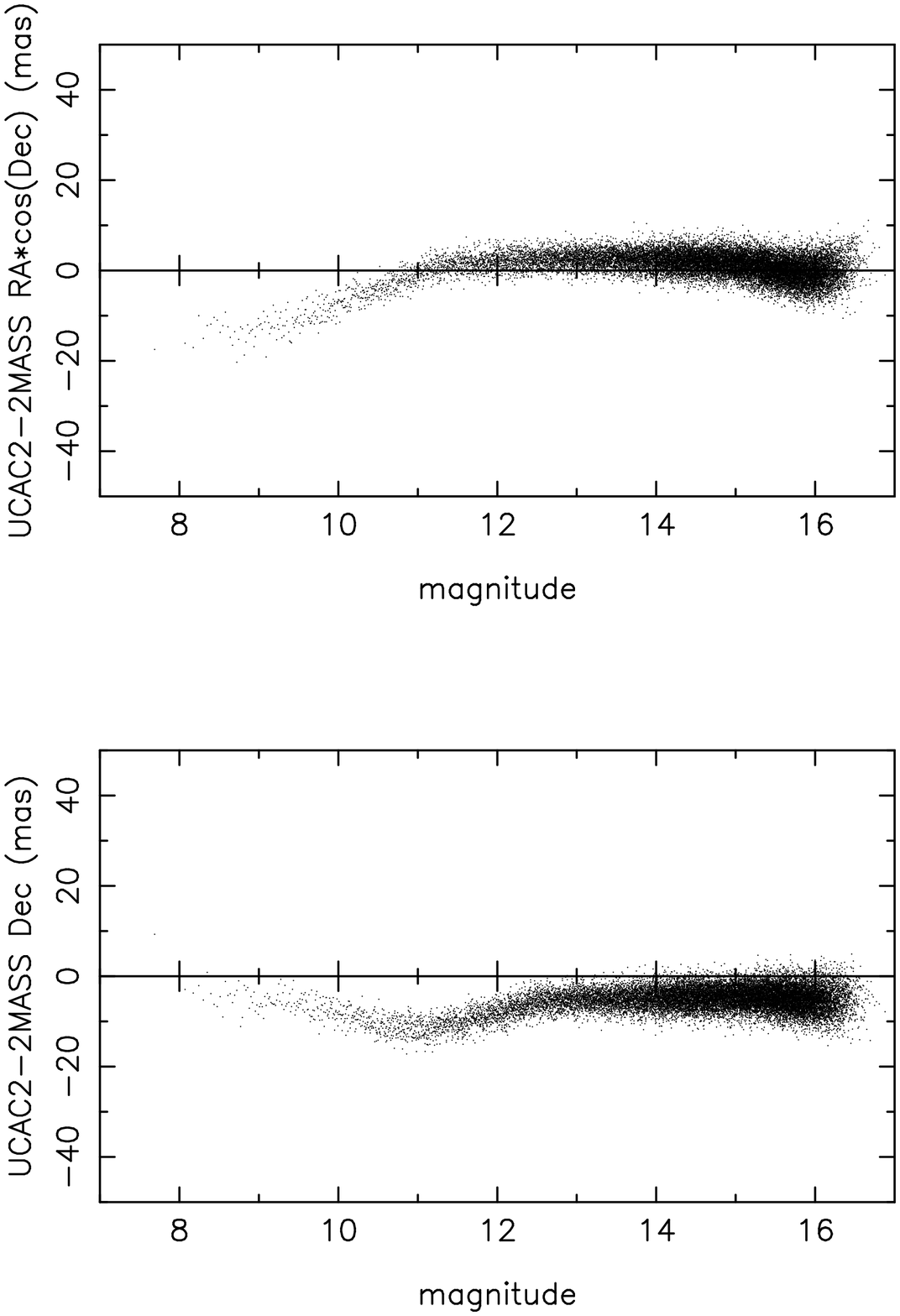}
\caption{Position differences UCAC2 $-$ 2MASS at the 2MASS epoch 
  using UCAC2 proper motions, as function of magnitude.
  These data are for the southern hemisphere.}
\end{figure}

\begin{figure}
\includegraphics[scale=.50]{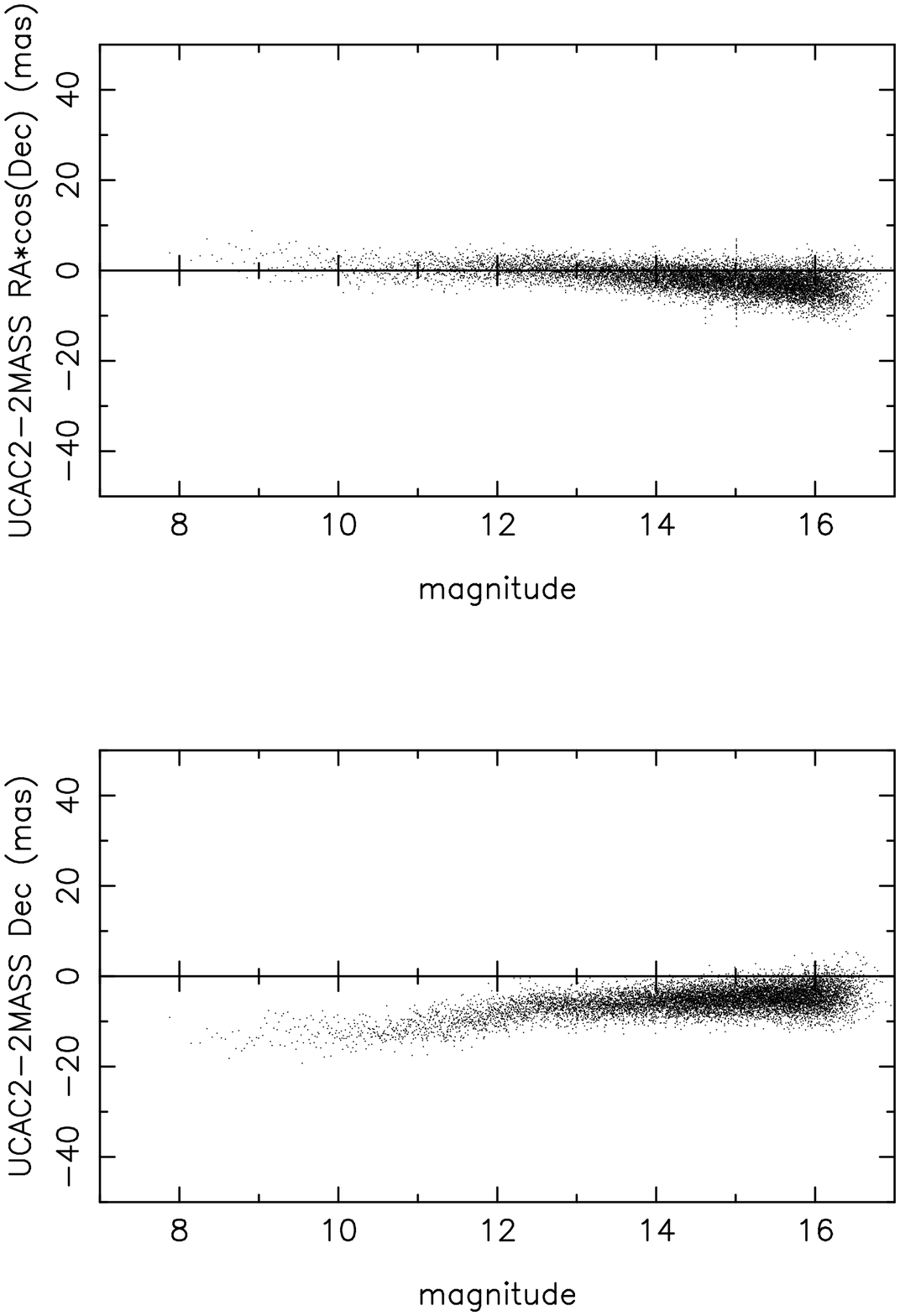}
\caption{Similar to the previous figure but for the northern hemisphere.}
\end{figure}

\clearpage

\begin{figure}
\includegraphics[scale=.50]{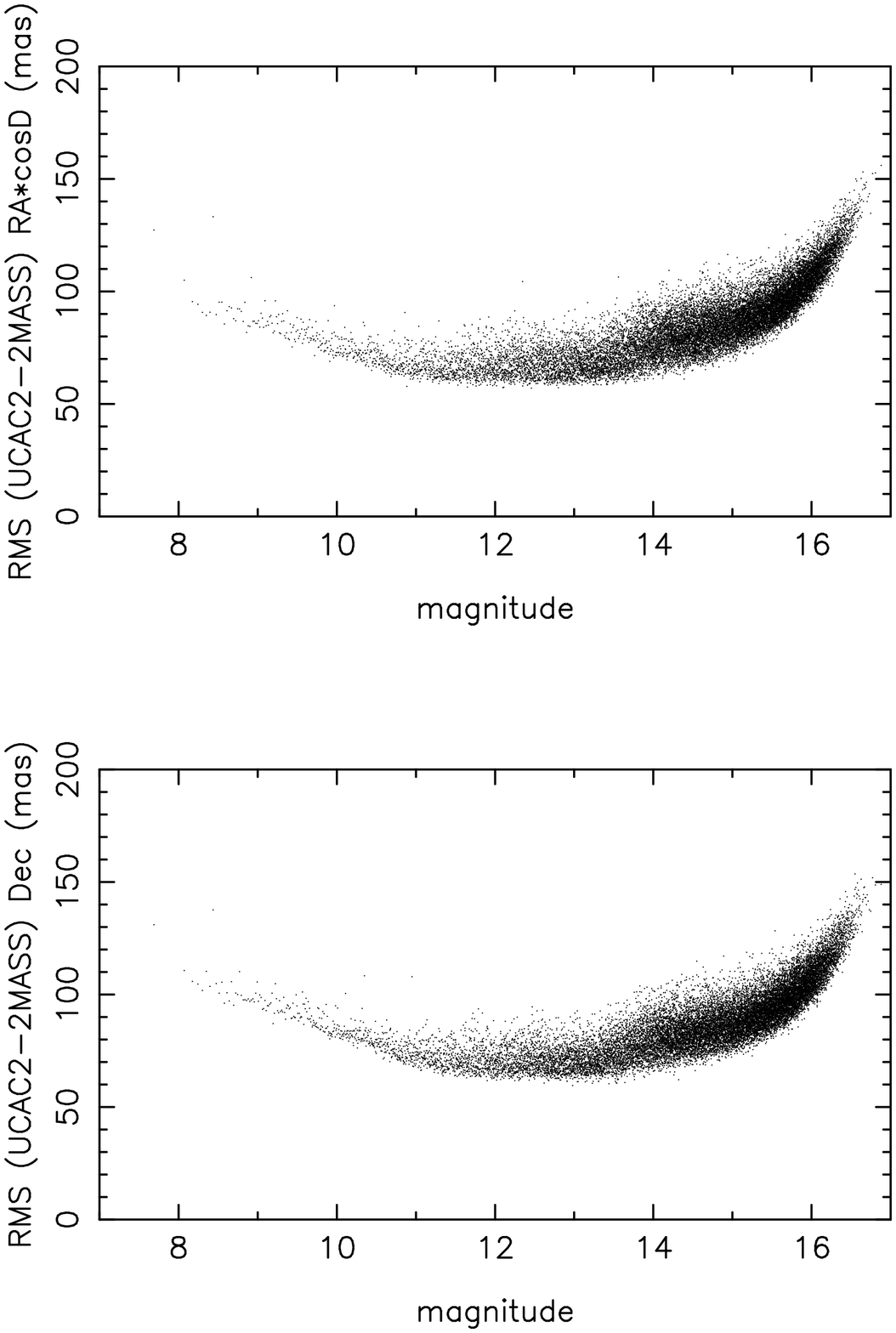}
\caption{RMS position differences UCAC2 $-$ 2MASS at the 2MASS epoch 
  using UCAC2 proper motions, as function of magnitude.
  These data are for the southern hemisphere.}
\end{figure}

\begin{figure}
\includegraphics[scale=.50]{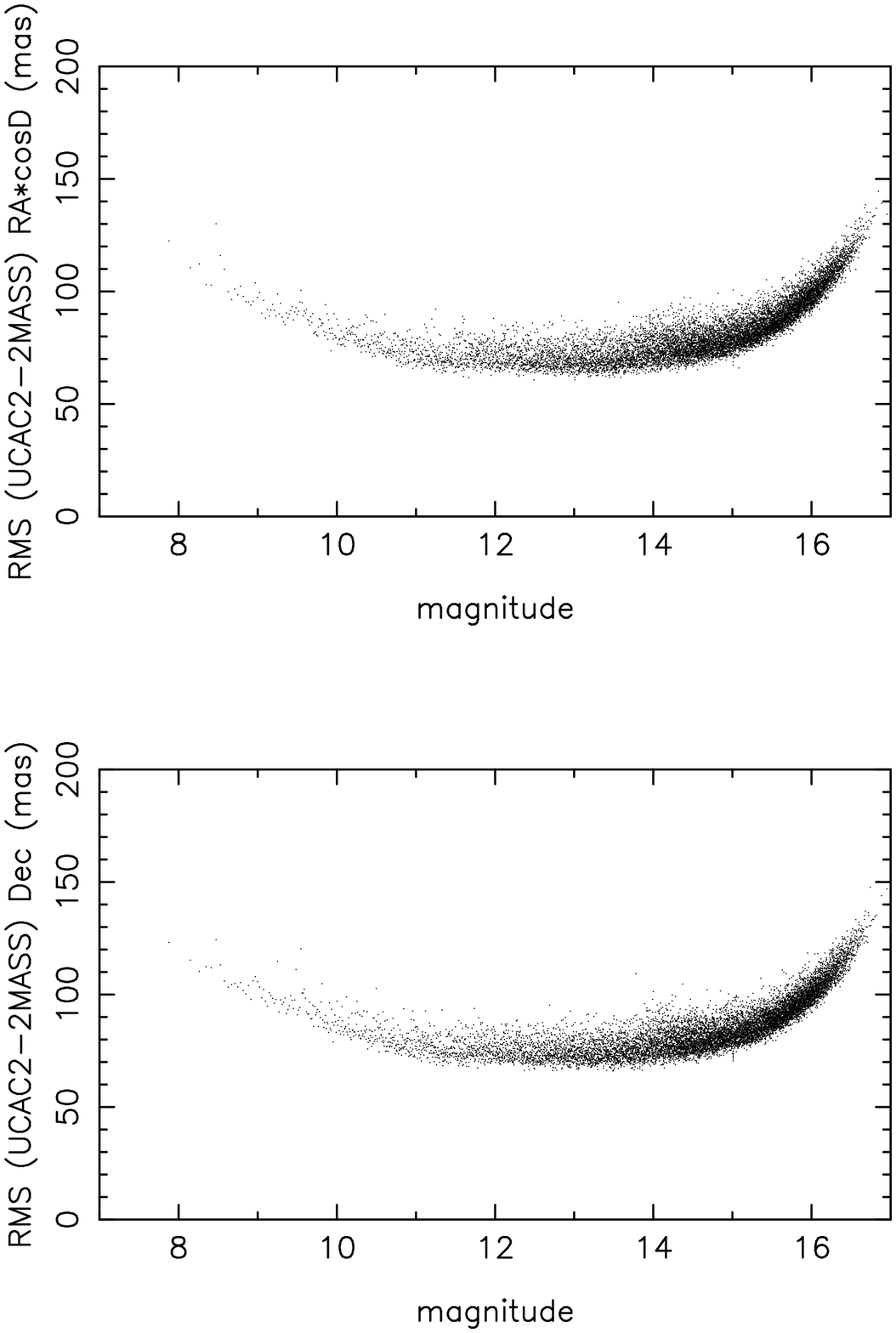}
\caption{Similar to the previous figure but for the northern hemisphere.}
\end{figure}






\clearpage

\begin{table}
\begin{center}
\caption{Number of stars in UCAC3 common with other catalogs or data sets.}
\vspace*{5mm}
\begin{tabular}{rrl}  
\tableline\tableline
 number  & catalog & catalog or data set name \\
of stars &   flag  &   \\
\tableline
      65,392  &  1  & Hipparcos \\
    2,386,607 &  2  & Tycho-2 \\
    4,098,873 &  3  & AC2000 \\
      270,823 &  4  & AGK2 Bonn \\
      960,074 &  5  & AGK2 Hamburg \\
    4,320,925 &  6  & Hamburg Zone Astrograph \\
    2,970,383 &  7  & USNO Black Birch Astrograph, Yellow lens\\
    1,043,857 &  8  & Lick Observatory 50cm Astrograph \\
   85,563,642 &  9  & SuperCOSMOS data\\
   51,112,855 & 10  & SPM Yale/San-Juan catalog (YSJ1) \\
       51,297 &  -  & high proper motion stars from 
         external catalogs\tablenotemark{a} \\
   98,114,307 &  -  & 2MASS\tablenotemark{b} \\
  100,766,420 &  -  & total number of entries in UCAC3 \\
\tableline
\end{tabular}
\tablenotetext{a}{identified by MPOS star number (last column in
  catalog data records, see also text) over 140,000,000}
\tablenotetext{b}{identified by separate 2MASS star identifier flag}
\end{center}
\end{table}


\begin{table}
\begin{center}
\caption{Adopted systematic errors which are added (RMS) to
         internal, random errors of star positions to obtain
         realistic weights before calculating proper motions.}
\vspace*{5mm}
\begin{tabular}{rl}  
\tableline\tableline
 error  & catalog name  \\
 (mas)  &   \\
\tableline
   1 & Hipparcos \\
  10 & Tycho-2   \\
  70 & AC2000.2   \\
  30 & AGK2 Bonn     \\
  30 & AGK2 Hamburg     \\
  20 & ZA  \\
  20 & BY  \\
  15 & Lick Astrograph\\
 100 & SuperCOSMOS    \\
  10 & SPM  \\
   5 & UCAC mean CCD position \\
  80 & all others \\
\tableline
\end{tabular}
\end{center}
\end{table}

\clearpage

\begin{deluxetable}{rlcllr}
\tabletypesize{\scriptsize}
\tablecaption{Data items for each star in UCAC3}
\tablewidth{0pt}
\tablehead{
\colhead{item} & \colhead{label} & \colhead{format}\tablenotemark{a} & 
\colhead{unit} & \colhead{description} & \colhead{remark} 
}
\startdata
 1& ra    &I*4&mas       & right ascension at  epoch J2000.0 (ICRS)& (1)\\
 2& spd   &I*4&mas       & south pole distance epoch J2000.0 (ICRS)& (1)\\
 3& im1   &I*2&millimag  & UCAC fit model magnitude                & (2)\\
 4& im2   &I*2&millimag  & UCAC aperture  magnitude                & (2)\\
 5& sigmag&I*2&millimag  & UCAC error on magnitude                 & (3)\\
 6& objt  &I*1&          & object type                             & (4)\\
 7& dsf   &I*1&          & double star flag                        & (5)\\
 8& sigra &I*2&mas       & s.e. at central epoch in RA (*cos Dec)  &  \\
 9& sigdc &I*2&mas       & s.e. at central epoch in Dec            &  \\
10& na1   &I*1&          & total numb. of CCD images of this star  &  \\
11& nu1   &I*1&          & numb. of CCD images used for this star  & (6)\\
12& us1   &I*1&          & numb. catalogs (epochs) used for proper motions& \\
13& cn1   &I*1&          & total numb. catalogs (epochs) initial match& \\
14& cepra &I*2&0.01 yr   & central epoch for mean RA, minus 1900   &  \\
15& cepdc &I*2&0.01 yr   & central epoch for mean Dec,minus 1900   &  \\
16& pmrac &I*4&0.1 mas/yr& proper motion in RA*cos(Dec)            &  \\
17& pmdc  &I*4&0.1 mas/yr& proper motion in Dec                    &  \\
18& sigpmr&I*2&0.1 mas/yr& s.e. of pmRA * cos(Dec)                 &  \\
19& sigpmd&I*2&0.1 mas/yr& s.e. of pmDec                           &  \\
20& id2m  &I*4&          & 2MASS pts key star identifier           &  \\
21& jmag  &I*2&millimag  & 2MASS J  magnitude                      &  \\
22& hmag  &I*2&millimag  & 2MASS H  magnitude                      &  \\
23& kmag  &I*2&millimag  & 2MASS Ks magnitude                      &  \\
24& icqflg&I*1&(3 items) & 2MASS cc.flg*10 + phot.qual.flag, J,H,Ks& (7)\\
25& e2mpho&I*1&(3 items) & 2MASS error photom. (1/100 mag), J,H,Ks & (8)\\
26& smB   &I*2&millimag  & SuperCOSMOS (SC) Bmag                   &  \\
27& smR2  &I*2&millimag  & SC R2mag                                & (9)\\
28& smI   &I*2&millimag  & SC Imag                                 &  \\
29& clbl  &I*1&          & SC star/galaxy classif./quality flag    &(10)\\
30& qfB   &I*1&          & SC quality flag Bmag                    &(11)\\
31& qfR2  &I*1&          & SC quality flag R2mag                   &(11)\\
32& qfI   &I*1&          & SC quality flag Imag                    &(11)\\
33& catflg&I*1&(10 items)& mmf flag for 10 major catalogs matched  &(12)\\
34& g1    &I*1&          & Yale SPM object type (g-flag)           &(13)\\
35& c1    &I*1&          & Yale SPM input cat.  (c-flag)           &(14)\\
36& leda  &I*1&          & LEDA galaxy match flag                  &(15)\\
37& x2m   &I*1&          & 2MASS extend.source flag                &(16)\\
38& rn    &I*4&          & MPOS star number; identifies high PM stars&(17)\\
\enddata
\tablenotetext{a}{ I means integer, followed by the number of bytes.}
\tablecomments{The extensive remarks are given in the readme file of UCAC3.}
\end{deluxetable}

\end{document}